 \documentclass[twocolumn]{aastex631}

\shorttitle{CN Light Curve}
\shortauthors{Kato et al.}

\begin{document}

\title{
Fully self-consistent nova explosion models reproducing
light curves of KT Eri, V339 Del, V597 Pup, and SMC NOVA 2016-10a}


\author[0000-0002-8522-8033]{Mariko Kato}
\affil{Department of Astronomy, Keio University,
Hiyoshi, Kouhoku-ku, Yokohama 223-8521, Japan}
\email{mariko.kato@hc.st.keio.ac.jp}

\author[0000-0002-0884-7404]{Izumi Hachisu}
\affil{Department of Earth Science and Astronomy,
College of Arts and Sciences, The University of Tokyo,
3-8-1 Komaba, Meguro-ku, Tokyo 153-8902, Japan}
%\email{izumi.hachisu@outlook.jp}

\author{Hideyuki Saio}
\affil{Astronomical Institute, Graduate School of Science,
 Tohoku University, Sendai 980-8578, Japan}
% \email{saio@astr.tohoku.ac.jp}

%\and


%\altaffiltext{1}{Visiting Astronomer, Cerro Tololo Inter-American Observatory\
.
%CTIO is operated by AURA, Inc.\ under contract to the National Science
%\altaffiltext{2}{Society of Fellows, Harvard University.}
%\altaffiltext{3}{present address: Center for Astrophysics,
%    60 Garden Street, Cambridge, MA 02138}
%\altaffiltext{4}{Visiting Programmer, Space Telescope Science Institute}
%\altaffiltext{5}{Patron, Alonso's Bar and Grill}


\begin{abstract}
The rising phase toward the optical maximum of a classical nova 
is one of the last frontiers of nova study.  Constructing 
free-free emission model light curves based on our fully self-consistent
nova explosion models, we present several theoretical light curves of
classical novae and compare them with the four novae having the observed
rising phase toward the optical maximum. 
Our 1.25 $M_\sun$ white dwarf (WD) models show excellent agreements with
the light curves of KT Eri, V339 Del, and V597 Pup while our 1.35 $M_\sun$
WD models are consistent with the light curves of SMC NOVA 2016-10a.
These agreements indicate that the light curves toward the optical 
maximum of these novae are dominated by free-free emission, rather than 
by photospheric emission. 
Our results justify the previously obtained WD masses and 
distance moduli for these novae, and shows that the post-maximum evolution 
can be well approximated with the evolution sequences of steady-state 
envelope solutions. 
\end{abstract}




%%%%%%%%%%%%%%%%%%%%%%%%%%%%%

\section{Introduction}
\label{introduction}

A nova is a thermonuclear runaway event that occurs on an accreting 
white dwarf (WD) in a binary \citep[e.g.,][]{nar80, ibe82, 
pri95k, sio79, spa78,taj15sn}. 
After the thermonuclear runaway sets in, the WD becomes 
bright and its surface temperature increases to  
$T_{\rm ph} \sim 10^6$ K and emits supersoft X-rays in a short time.
This very early phase phenomenon is called the X-ray flash  
\citep{kon22wa,kat22shapjl}. 
The WD envelope expands and the photospheric temperature decreases.
Strong winds emerge from the WD photosphere when the photospheric
temperature decreases to $T_{\rm ph} \sim 10^{5.4}$ K 
\citep[e.g.,][]{kat94h}.
Then the X-ray flash phase ends. 
Winds are accelerated deep inside the photosphere \citep{fri66, rug79} 
so that they are called the optically thick winds. 
Winds become optically thin outside the WD photosphere.
The WD photosphere further expands toward optical maximum 
and the nova becomes brighter in the optical band
owing to free-free emission from the optically thin ejecta
\citep[e.g.,][]{enn77, gal76, hac06kb, hac25kv1674her2}. 

At the optical maximum, the WD photosphere reaches maximum (radius) 
and the wind mass-loss rate also reaches maximum. 
After the maximum expansion of the photosphere,
the WD envelope (interior to the photosphere) quickly loses its mass 
mainly due to wind mass loss. The mass loss rate becomes smaller 
and thus, free-free emission becomes fainter.  
As most of the envelope mass is blown in the wind,
the WD photosphere shrinks and the photospheric temperature increases. 
Then the optically thick winds finally end and 
the nova enters the supersoft X-ray source (SSS) phase
\citep{kra96, sch11, hen11, hen15, hen18, pag22}.  
Finally, nuclear burning ends and the WD becomes faint. 

% This is the main physical process that occurs on the WD. 

Nova light curves have been difficult to calculate with 
Henyey-type evolution codes especially after the opacity tables
are revised \citep{igl93, igl96, sea94}
that shows a large Fe peak at around $\log T {\rm ~(K)}=5.2$ which accelerates 
envelope expansion. The numerical difficulty appears
when the diffusive luminosity approaches the local Eddington luminosity
in the extended nova envelope. 
Many authors adopt some technique 
%to skip the extended stage 
to continue evolution calculation, i.e., to skip the extended stage assuming 
a large mass-loss \citep[e.g., ][]{den13hb, ma13, wu17, chen19}  
or including convection in the supersonic region \citep[e.g., ][]{she22}. 
If we adopt such methods, we can hardly calculate accurate 
optical light curves because the photospheric temperature becomes $\sim 10^4$
K or less at the optical peak, much lower than the Fe opacity peak 
($\log T$ (K)$\approx 5.2$)
\citep[see][for a review and comparison
among various approximations of the numerical methods]{kat17palermo}.
 
Thus, actually no accurate theoretical light curves have ever been presented
except Kato's group.
A rare exception is \citet{hil14} who presented theoretical light
curves of nova outbursts based on the calculation by \citet{pri95k}. 
However, their $V$ light curves show a flat-peak followed by a sudden drop  
as a rectangular shape, which are very different from gradual declines 
in observed nova light curves.  The reason is partly a small number of
mass grids \citep[see Figure 9 in][]{hil16}
and inadequate surface boundary condition 
as pointed out in Figure 6 of \citet{kat17palermo}.

\citet{kat22sha} succeeded, for the first time, 
in calculating one cycle of nova evolution 
including extended phase of the envelope. They determined wind mass loss rates
which are consistent with the continuum-radiation driven wind by the Fe peak.
Using this method, \citet{kat25v1674her} calculated the $V$ light curve 
of a 1.35 $M_\sun$ WD and reproduced the V1674 Her observation
including the very early rising phase \citep[see also ][for an extremely
early phase]{hac25kv1674her2}. 

The optical rising phase toward the optical peak of a nova outburst 
has been poorly studied partly because of lack of theoretical models 
as well as the paucity of observational data.  
We have calculated several nova light curves of 1.25 $M_\sun$
and 1.35 $M_\sun$ WDs, which enable us quantitative comparison 
with observed nova outburst data around the optical peak. 
Our aim of this paper is to 
% present our light curve models around the peak 
compare the observed data with our theoretical light curves 
and clarify its physics. 

This paper is organized as follows. Section \ref{sec_model} explains 
our numerical method and presents our optical light 
curves based on our free-free emission model. 
Section \ref{sec_obs} compares our model light curves with the
four classical novae, KT Eri, V339 Del, V597 Pup, and SMC Nova 2016-10a.
Our discussion and conclusions follow in 
Sections \ref{sec_discussion} and \ref{sec_conclusion}.

% Table.1
%\startlongtable
%\begin{deluxetable*}{llllllll}
\begin{deluxetable*}{lllllrlllllccll}
\tabletypesize{\scriptsize}
\tablecaption{Model Properties\tablenotemark{a}
\label{table_models}}
%\tablewidth{0pt}
\tablehead{\colhead{Model} &\colhead{$M_{\rm WD}$} &
 \colhead{$\dot M_{\rm acc}$} &
\colhead{C mix} &
\colhead{$\log T_{\rm WD}$} &
\colhead{$t_{\rm rec}$} &
\colhead{$L_{\rm nuc}^{\rm max}$} & \colhead{$M_{\rm ig}$}&
\colhead{$\log T^{\rm max}$}& 
%\colhead{$t_{\rm flash}$\tablenotemark{a}}\\
\colhead{$R_{\rm ph}$}&
\colhead{$-\dot M_{\rm wind}^{\rm max}$}\\ 
\colhead{ }&
\colhead{ ($M_\sun$) } &
\colhead{($M_\sun$ yr$^{-1}$)   } &
\colhead{ }&
\colhead{(K) }&
\colhead{(yr)}&
\colhead{($ 10^9 L_\sun $) } &
\colhead{($10^{-6}M_\sun$) } &
 \colhead{(K) }&
 \colhead{($R_\sun$) }&
\colhead{($10^{-5}M_\sun$ yr$^{-1}$) }  
%\colhead{ }
}
\startdata
%M125.3E-11&1.25 & $3\times 10^{-11}$& 0.1  & 7.42 &190,000  & 26.0  & 6.0 & 8.27 & $< $ hr\\ 
M125ac5E-11&1.25 & $5\times 10^{-11}$ & 0.1  & 7.44  &120,000 & 27.0  & 6.1 & 8.27 & 82 & 26 \\
M125ac1E-10&1.25 & $1\times 10^{-10}$ & 0.1  & 7.49  & 51,000 & 18.0  & 5.4 & 8.26 & 63 & 22 \\ 
M125ac5E-10&1.25 & $5\times 10^{-10}$ & 0.1  & 7.58   & 8,100 & 9.3  & 4.3  & 8.24 & 42 & 17 \\ 
M125ac1E-9&1.25 & $1\times 10^{-9}$   & 0.1  & 7.63  & 4,000  & 7.0  & 4.2  & 8.23 & 33 & 14 \\ 
M125ac5E-9&1.25 & $5\times 10^{-9}$   & 0.1  & 7.74   & 520   & 1.6  & 2.8  & 8.19 & 16 & 8.3  \\ 
\hline
M135ac1E-11&1.35 & $1\times 10^{-11}$ & 0.1 & 7.39  &160,000  & 31   & 1.6  & 8.33 & 21  & 12 \\ 
M135ac5E-10&1.35 & $5\times 10^{-10}$ & 0.1 & 7.61  &1,900    & 5.8  & 1.0  & 8.28 & 14  & 8.8 \\
M135ac5E-9&1.35 & $5\times 10^{-9}$   & 0.1 & 7.79  &  120    & 1.2  & 0.70 & 8.24 & 3.2 & 2.3  % (50)66(53)40 
\enddata
\tablenotetext{a}{Our control parameters
in our model calculations are WD mass $M_{\rm WD}$,
mass accretion rate $\dot{M}_{\rm acc}$,
and carbon enhancement C$_{\rm mix}$. The other quantities are the results 
of our evolution calculations.}
\end{deluxetable*}

\section{Model light curves of novae} \label{sec_model}

\subsection{Numerical Method} \label{sec_method}

We have calculated nova outburst cycles for 1.25 $M_\sun$ 
and 1.35 $M_\sun$ WDs with several mass accretion rates. 
We assume spherical symmetry and solve interior structure
from the center of the WD  up to the photosphere.  
We do not calculate
outside the WD photosphere, even if winds extend over the WD photosphere. 
We use a Henyey-type stellar evolution code consistently connected
with steady-state wind solutions in each time-step to determine
the wind mass-loss rate.  
The typical number of the mass zones is slightly less than or about 1000 
for the WD interior the region of which we solve with the Henyey code,  
and several thousands mass zones for the wind solutions which are
solved with the steady-state hydrodynamic code \citep{kat94h}. 
The time-steps, which depend on the mass-accretion rate, 
are taken to be 0.01-0.05 s during the thermonuclear runaway phase,
100-10,000 s near the optical peak, and a few tens of years 
for the quiescent phase.
This method requires complicated 
and time-consuming human iteration procedure to fit the interior
with the exterior envelope solution in each time step, 
but enable us to calculate one full cycle nova evolution including 
extended envelope phase around the optical peak. 

For the initial WD model 
% of our evolution calculations,
we have adopted a steady-state equilibrium model \citep{kat14shn}, in which
the heating by the mass accretion and nuclear energy generation is balanced
with the cooling by radiative diffusion and neutrino emission.
Starting from the steady-state WD model,
the flash properties reach a limit cycle before 5th flash cycle,
which indicates that the interior structure of our initial model is
already a good approximation to the structure reached after many flashes.
The method of our numerical calculation is described
in \citet{kat22sha, kat24M1213, kat25v1674her}.

We have assumed a solar composition for accreting matter
($X=0.7$, $Y=0.28$, and $Z=0.02$). 
The composition of the CO core is composed of
48\% of $^{12}$C, 50\% of $^{16}$O, and 2\% of $^{20}$Ne by mass.
This ratio of C, O, and Ne does not affect our results of a nova outburst. 
We used the OPAL opacity tables \citep{igl96}.

Many classical novae show heavy element enrichment
in ejecta \citep[e.g.,][]{van96,van97,geh98tw,hac06kb,sch07,sok23}. 
%Also V1674 Her is suggested non-solar heavy element enrichment
%in the ejecta \citep{,hab24}. 
% (see Section \ref{sec_composition} for detail).
%
To mimic such a heavy element enrichment, 
we have increased carbon mass fraction in the
hydrogen-rich envelope by 0.1
and decreased helium mass fraction by the same amount at the
beginning of thermonuclear runaway \citep{sta20,kat24M1213}. 
The amount of carbon enrichment does not affect the results
because the energy generation at ignition, $L_{\rm nuc}^{\rm max}$, 
is unchanged between the carbon mixtures of 0.1 and 0.2 \citep{kat22shc}.

The model parameters and characteristic physical values 
are listed in Table \ref{table_models}.
From left to right, model name, the WD mass, mass accretion rate
in quiescent phase, assumed carbon enhancement in the hydrogen-rich envelope,
temperature at the WD center, recurrence period of nova outbursts,
maximum nuclear burning energy generation rate, ignition mass, i.e.,
mass of the hydrogen-rich envelope at the onset of thermonuclear runaway,
the maximum temperature at hydrogen burning, 
photospheric radius at the maximum expansion, i.e., 
the WD radius at the optical peak, and 
the maximum mass-loss rate which is reached at the maximum expansion
of the photosphere.  

The basic properties of one cycle of a nova outburst, 
internal structures, optical and SSS light curves are already 
described in \citet{kat25v1674her} for models M135ac1E-11 and M135ac5E-10.
The data of model M135ac5E-9 are partly reported in \citet{kat22shapjl}
only for the X-ray flash phase. 
The 1.25 $M_\sun$ WD models show essentially the same 
properties as those of the 1.35 $M_\sun$ WD models.  
Thus, in the present paper we concentrate on the early outburst phase, 
that is, the optical rising phase, optical peak, and early decay phase.

%%%%%%%%%%%%%%%%%%%%%%%%%%%%%

% 1.35 $M_\sun$ with $5.0 \times 10^{-10}~M_\sun$ yr$^{-1}$

% universal decline law ($L_\nu \propto t^{-1.75}$) 

\subsection{Optical Light Curve}
\label{optical_light_curve}

% Fig.1
\begin{figure}
%\epsscale{0.7}
\epsscale{1.15}
%\plotone{all.ps}
\plotone{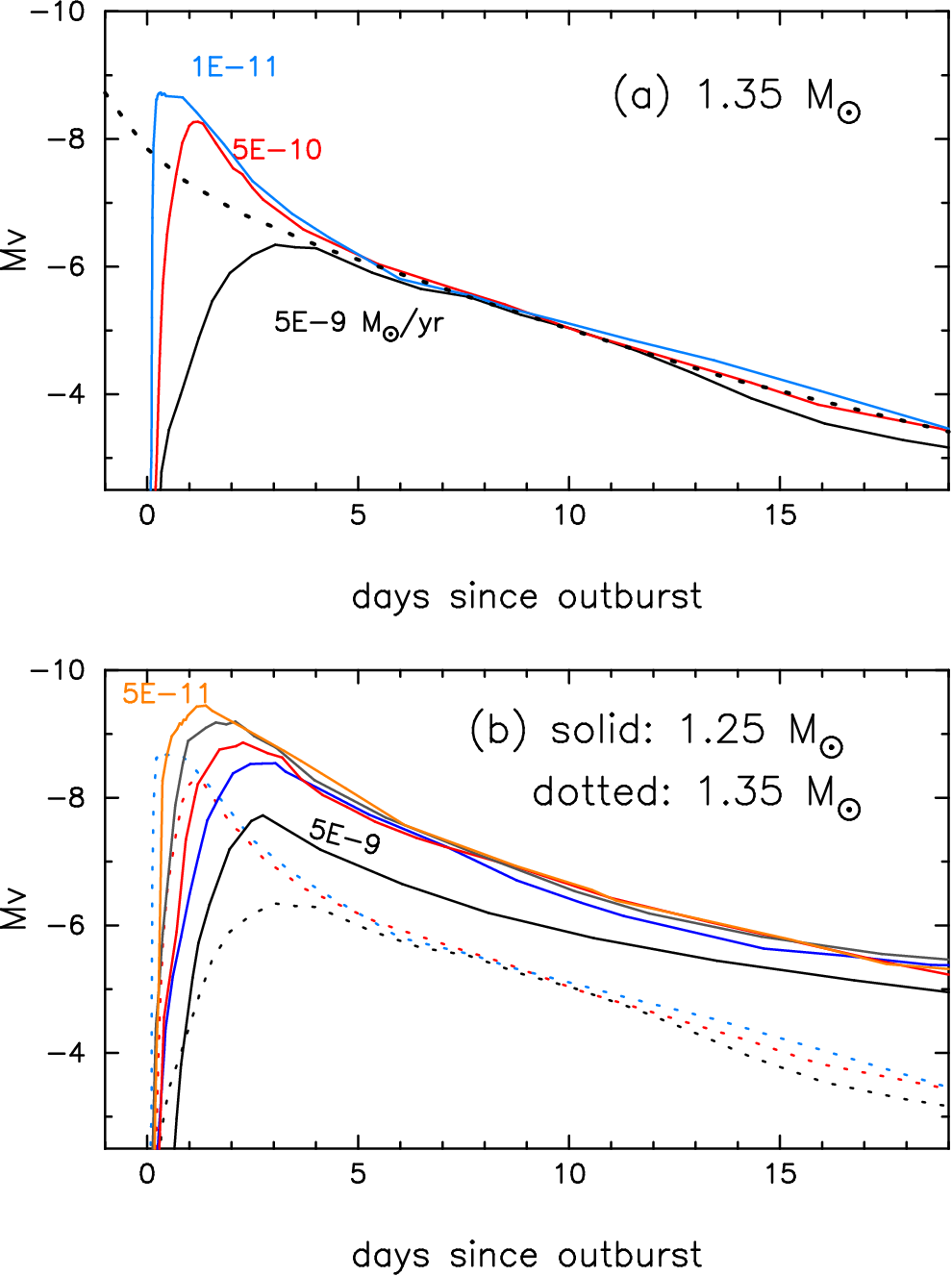}
\caption{Theoretical model light curves for novae. 
(a) Our 1.35 $M_\sun$ WD model with mass-accretion rates of 
$\dot M_{\rm acc}=1\times 10^{-11}~M_\sun$ yr$^{-1}$ (cyan-blue line),
$5\times 10^{-10}~M_\sun$ yr$^{-1}$ (red line), and 
$5\times 10^{-9}~M_\sun$ yr$^{-1}$ (black line).  The origin of time
($t=0$) is set at the onset of thermonuclear runaway.
The black dotted line indicates the 
sequence of steady-state wind solutions \citep{kat94h}  
for a 1.35 $M_\sun$ WD with the envelope chemical composition of $X=0.55$, 
$Y=0.3$, $Z=0.02$, $X_{\rm C}=0.03$ and $X_{\rm O}=0.1$. 
The radius at the bottom of the envelope (i.e., nuclear burning region)
is assumed to be $\log R/R_\sun=-2.48$, which is taken from model M135ac5E-10. 
(b) Same as panel (a), but we added 1.25 $M_\sun$ WD models. 
The same color indicates the same accretion rate as that in panel (a). 
Orange line: $\dot M_{\rm acc}=5\times 10^{-11}~M_\sun$ yr$^{-1}$. 
Gray line: $1\times 10^{-10}~M_\sun$ yr$^{-1}$.
Blue line: $1\times 10^{-9}~M_\sun$ yr$^{-1}$. 
}\label{all}
% source: rn/m125.5e-9.C10.atL3/all.wip
\end{figure}

%$1.35 ~M_\sun$ WD with the mass accretion rate
%of $\dot M_{\rm acc}=5\times 10^{-10}~M_\sun$ yr$^{-1}$ (Model B). 
%These models are partly presented by \citet{kat22shapjl, kat22shc} 
%only for the X-ray flash.  

Assuming that the $V$ band luminosity is dominated by free-free 
emission \citep[e.g.,][]{enn77, gal76}, we have calculated optical
$V$ light curves of each model.  In the present paper, 
we do not include the contribution of the WD photosphere 
(blackbody assumption) because it is much fainter than that of
free-free emission (see Discussion Section \ref{dis_bb}).
The $V$ band luminosity of free-free emission from optically thin 
plasma outside the WD photosphere can be simplified as
\citep{hac06kb, hac20skhs}
\begin{equation}
L_{V, \rm ff,wind} = A_{\rm ff} ~{{\dot M^2_{\rm wind}}
\over{v^2_{\rm ph} R_{\rm ph}}},
\label{free-free_flux_v-band}
\end{equation}
where $\dot{M}_{\rm wind}$ is the wind mass loss rate,
$v_{\rm ph}$ the wind velocity at the WD photosphere,
and $R_{\rm ph}$ the photospheric radius of the WD. 
%We use the wind mass loss rate $\dot{M}_{\rm wind}$, 
%the velocity at the photosphere $v_{\rm ph}$, 
%and the photospheric radius $R_{\rm ph}$ calculated for
%our $1.35 ~M_\sun$ WD model.  
The coefficient $A_{\rm ff}$ depends on the WD mass and chemical 
composition of the envelope.  We have already determined 
the coefficient $A_{\rm ff}$ by novae with known distances 
\citep{hac10k, hac16k, hac18kb, hac20skhs}. 
%The resultant light curves are consistent with each other.
We choose $A_{\rm ff}$ for the 1.35 $M_\sun$ models to match that of the
existing steady-state envelope model
that has the closest chemical composition.  
The resultant light curves are very consistent with 
V1674 Her \citep[see ][for more details]{kat25v1674her}.

For the 1.25 $M_\sun$ models, we have no corresponding steady-state envelope
models with the same chemical composition.  
We determined $A_{\rm ff}$ from an interpolation of 
steady-state envelope models in Tables 2 and 3 of \citet{hac25kv392per}.

% The two 1.35 $M_\sun$ models (M135ac1E-11, M135ac5E-10) are already 
%presented from the beginning of the outburst until the end of SSS phase 
% \citep{kat25v1674her}. 
% Evolution of 1.25 $M_\sun$ models are essentially the same. 
% Thus, the present paper concentrates on the early evolution 
% including the optical peak.
 
Figure \ref{all} shows our theoretical light curves. 
% Characteristic properties of these models are explained with 
% the difference of the model parameters. 
%
Panel (a) shows three 1.35 $M_\sun$ WD models. 
Among them, the lowest mass-accretion rate model of  M135ac1E-11 shows
the brightest peak.  This model has the largest ignition mass
among the three models (Table \ref{table_models}) so that the WD
photosphere extends to the largest size among the three models.
The wind mass-loss rate $-\dot M_{\rm wind}^{\rm max}$ is larger 
for a larger extended WD photosphere \citep{kat25v1674her}. 
The luminosity of free-free emission strongly depends on the wind
mass-loss rate (see Equation (\ref{free-free_flux_v-band})), so that
it attains the brighter optical maximum.  The highest mass-accretion rate
model M135ac5E-9 shows a slower rise toward the optical maximum 
because the ignition mass is smaller and this results in a
lower maximum temperature $T^{\rm max}$  
at the nuclear burning region and a lower maximum nuclear 
burning rate $L_{\rm nuc}^{\rm max}$ \citep[see, e.g., ][for more
details]{kat24M1213}.

After day 4, the three model light curves converge into one.  
The dotted black line indicates the light curve for the sequence
of steady-state envelope solutions which we call the steady state 
(light curve) model \citep{hac06kb}.
This indicates that the envelope structure 
settled down to a thermal equilibrium which is independent of 
the initial condition.   

Figure \ref{all}(b) shows five 1.25 $M_\sun$ WD models (solid lines)
in addition to the 1.35 $M_\sun$ WD models (dotted lines).  
The peak magnitudes of the 1.25 $M_\sun$ WD models are 
systematically higher than the 
1.35 $M_\odot$ models because the ignition masses are larger 
and then the wind mass loss rate reaches a larger value. 
This tendency has been reported in Figure 6 of \citet{hac20skhs}: 
for $M_{\rm WD} \gtrsim$ 1.1 $M_\sun$, the peak magnitude is fainter
in a more massive WD with the same mass-accretion rate. 

The five light curve models 
do not converge in this figure, but if we shift these models 
in the horizontal direction all the lines converge into one as shown 
later in Figure \ref{kteri}a.

\section{Comparison of model light curves with four novae} \label{sec_obs}

% Table.2
%\startlongtable
\begin{deluxetable*}{llllllll}
%\begin{deluxetable*}{llllllllllcclllll}
\tabletypesize{\scriptsize}
\tablecaption{Summary of four novae
\label{table_novae}}
%\tablewidth{0pt}
\tablehead{\colhead{Object} &
\colhead{Outburst}&
\colhead{$(m-M)_V$}&
\colhead{$E(B-V)$}&
\colhead{$d$} &
\colhead{$M_{\rm WD}$} &
\colhead{reference\tablenotemark{a}} \\
\colhead{ }&
\colhead{ year } &
\colhead{ }&
\colhead{ }&
\colhead{(kpc)   } &
\colhead{($M_\sun$) }&
\colhead{ }
}
\startdata
%V838 Her     & 1991 & 13.2--15.3 & 0.33--0.53 & 2.7--6.0 &$1.35\pm0.02$ &1,2\\
V597 Pup    & 2007 & $17.0\pm0.2$ & 0.3& $16 \pm2 $ &$1.2\pm0.05$ &1\\
KT Eri     & 2009 & $13.4\pm0.2$ & 0.08& $4.2 \pm0.4$ &$1.3\pm0.02$ &2\\
V339 Del   & 2013 & $12.2\pm0.2$ & 0.18& $2.1 \pm0.2$ &$1.25\pm0.05$ &3\\
SMCN 2016-10a & 2016 &$16.8 \pm0.2$ & 0.08 & $20.5 \pm2$  &$1.29\pm0.05$ &4\\
\enddata
\tablenotetext{a}{1: \citet{hac10k}, 2: \citet{hac25kw}, 
3: \citet{hac24km}, 4: \citet{hac18kb} }
%\tablenotetext{d}{Increased carbon mass-fraction by 0.1 at ignition.}
\end{deluxetable*}

In this section, we compare our theoretical light curves with the four 
classical novae listed in Table \ref{table_novae}
whose optical peak are clearly observed including before and after the peak. 
These novae have been analyzed by comparing with steady state sequence models
as well as by the time-stretching method \citep[e.g.,][]{hac18kb}. 
Their WD masses and distance moduli in the $V$ band $\mu_V\equiv (m-M)_V$
are well constrained. 
In the present paper, we concentrate on the first few tens of days 
from the outburst to avoid secondary effects such as nebular emission lines,
dust formation, and contributions from the optically-thin shocked shell.
These effects occur outside the WD photosphere 
and contribute later to the light curves as analyzed 
in the papers listed in Table \ref{table_novae}. 
We will summarize the effects of shock formation in 
Section \ref{dis_shock}. 

\subsection{KT Eri 2009} \label{sec_kteri}

% Fig.2
\begin{figure}
%\epsscale{0.7}
\epsscale{1.15}
%\plotone{kteri.f.ps}
\plotone{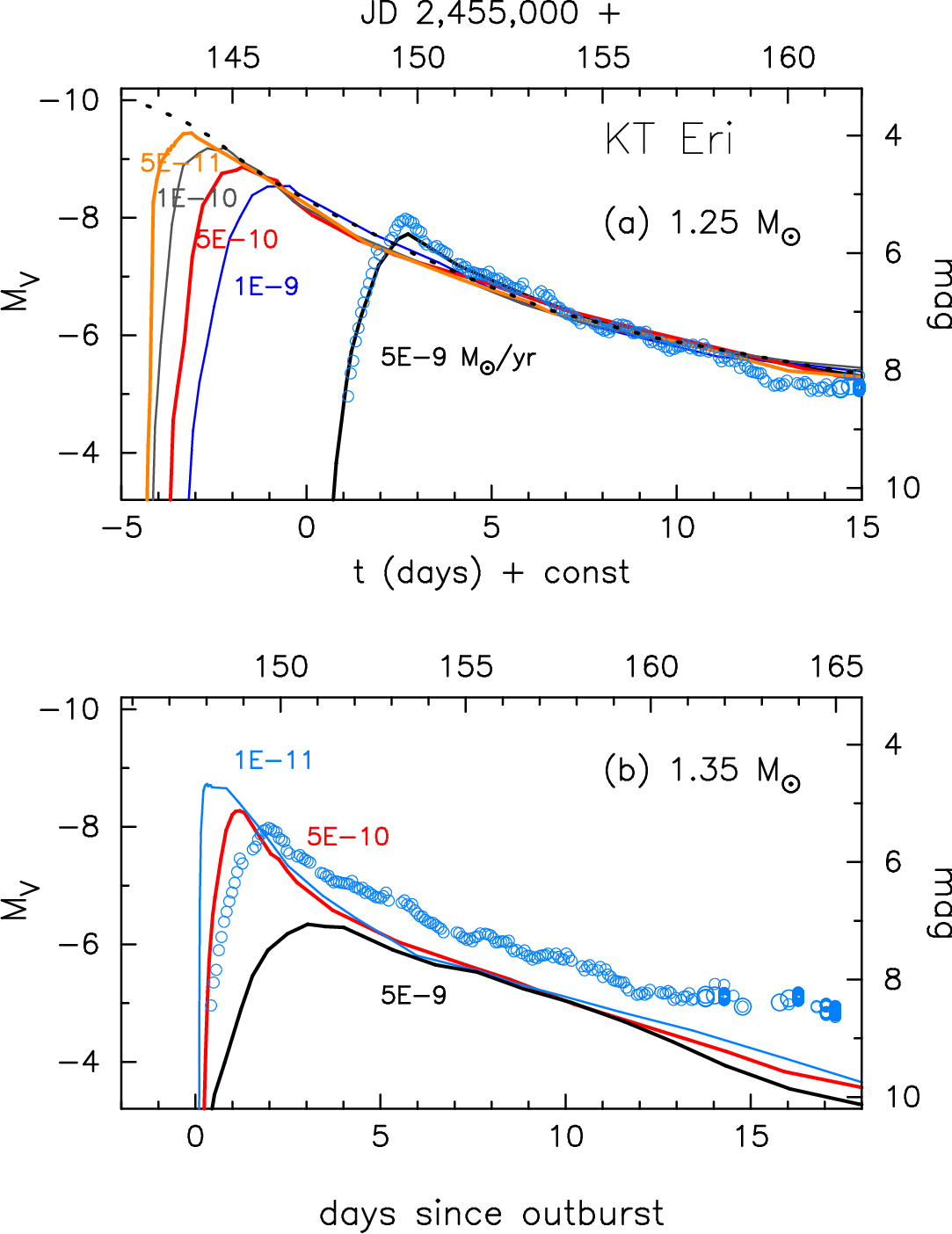}
\caption{Comparison between the KT Eri $V$/SMEI light curves and
our theoretical model light curves.
The cyan-blue circles denote the KT Eri data for the distance modulus in
the $V$ band $\mu_V\equiv (m-M)_V=13.4$ as listed in Table \ref{table_novae}. 
All the KT Eri data are the same as those in \citet{hac25kw}, i.e., 
SMEI (before JD 2455162), and VSOLJ and AAVSO (after JD 2455151).  
(a) Each 1.25 $M_\sun$ WD model is shifted horizontally to fit the 
post-maximum phase with that of KT Eri. 
%after the peak of $5\times 10^{-9}~M_\sun$ yr$^{-1}$ model.
Thus, their outburst day is not just at $t=0$ but shifted by $\Delta t$. 
KT Eri is well fitted to the line of 
$\dot M_{\rm acc}=5 \times 10^{-9}~M_\sun$ yr$^{-1}$. 
The sequence of steady-state envelope solutions (dotted black line)
is added for our 1.25 $M_\sun$ WD with the envelope chemical
composition of $X=0.6$, $Y=0.25$, $Z=0.02$, and $Z_{\rm CNO}=0.13$.
(b) Our 1.35 $M_\sun$ WD model light curves are depicted
instead of the 1.25 $M_\sun$ WD models. 
The onset of thermonuclear runaway is at $t=0$ for all models. 
%The open black circles indicate the KT Eri data shifted rightward
%by 0.7 days and downward by 0.8 mag.
%%%in the horizontal direction,  in the vertical direction.
% \citep[see for details][]{hac25kw}.
}\label{kteri}
\end{figure}

KT Eri is a classical nova, discovered on UT 2009 November 25 (JD 2,455,160.5)
at $V=8.1$ by K. Itagaki \citep{yam09ig}. 
The orbital period is obtained to be $P_{\rm orb}=2.6$ days
\citep{schaefer22wh}, which is exceptionally long for classical novae.

Figure \ref{kteri} shows the prediscovery light curve
constructed by \citet{hou10bh} 
based on the data of the Solar Mass Ejection Imager 
\citep[SMEI,][]{hou10bh}, which  
cover 3 magnitude rapid rise toward maximum followed by 
a much slower decay. The dense optical data around the optical peak enable us 
to accurately compare them with our theoretical light curves. 
% 8.44 to 5.4180  
In this figure we adopt the distance modulus in the $V$ band
of $\mu_V\equiv (m-M)_V =13.4$ 
after \citet{hac25kw} who obtained  $(m-M)_V =13.4 \pm 0.2$
with the Gaia distance $d=4.2 \pm 0.4$ kpc \citep{bai21gaia}
and the reddening of $E(B-V)=0.08\pm0.02$ \citep{gre19}.

Figure \ref{kteri} also shows our model light curves.  Figure \ref{kteri}(a) 
depicts the 1.25 $M_\sun$ WDs with five different mass accretion rates.
Among them, the model with the mass accretion rate of $\dot M_{\rm acc}=
5 \times 10^{-9}~M_\sun$ yr$^{-1}$ shows a good agreement with the KT Eri
light curve.  In the later phase of $t>13$ days,
the light curve data becomes slightly fainter 
than our theoretical models, which suggests that a WD mass is possibly 
more massive than 1.25 $M_\sun$.  

Figure \ref{kteri}(b) shows our model light curves of a 1.35 $M_\sun$ WD. 
The cyan-blue circles are plotted for the same distance modulus of
$(m-M)_V =13.4$ as in panel (a).
We choose its rising phase of KT Eri to match the red line 
($t=0$ is the onset of thermonuclear runaway).
In the rising phase the increasing rate is between the red line 
and black line, which is consistent with the peak magnitude, 
i.e., slightly fainter than the red line but brighter than 
the black line. After the peak,  
KT Eri (cyan-blue circles) shows a slower decline than these lines 
and much brighter than the 1.35 $M_\sun$ models 
after day 3. 

%If we could assume a smaller distance modulus by 0.8 mag, i.e., 
%$(m-M)_V =13.4 - 0.8= 12.6$, 
%%%%than $(m-M)_V =13.4$, 
%and shifted rightward by 0.7 days,
%the cyan-blue circles move to the black circles. 
%These black circles show a good agreement with our model light curves 
%after day 4. Comparing with the black line, the peak magnitude is 
%much brighter and the rising phase would be too fast. 
We thus conclude that the 1.35 $M_\sun$ WD models do not correctly
reproduce the KT Eri light curve. 
The slower decline of the cyan-blue circles in the post-maximum phase 
suggests that the WD mass is less massive 
than 1.35 $M_\sun$, that is, the more massive the WD mass is,
the faster/steeper the decay/slope of $V$ light curve is
\citep[e.g.,][]{hac06kb}. 

\citet{hac25kw} analyzed the KT Eri light curve
over the full timespan of the outburst, 
i.e., from its optical peak until the end of the supersoft X-ray phase.
Their light curve model is based on the 
steady-state sequence and included the 
contributions of an irradiated accretion disk and companion star 
which are effective in a later phase. 
They obtained the WD mass to be $1.3 \pm 0.02~M_\sun$, the mean mass
accretion rate of $\dot{M}_{\rm acc}\sim 1\times 10^{-9} ~M_\sun$ yr$^{-1}$
in quiescent phase and the recurrence time to be $t_{\rm rec}\sim 3000$ yr. 
The present light curve fitting is restricted to around the peak
but includes the pre-maximum phase which is not analyzed in \citet{hac25kw}. 
In spite of such differences, our light curve fitting 
in Figure \ref{kteri} is consistent with their mass estimate. 

KT Eri was once suggested to be a recurrent nova of the recurrence time
$t_{\rm rec}$$\sim40$--50 yr \citep[e.g.,][]{schaefer22wh},
but our results ($t_{\rm rec}=520$ yr for M125ac5E-9 in Table
\ref{table_models}) clearly contradict the possibility of
a recurrent nova ($t_{\rm rec}\lesssim 100$ yr).

Also a good agreement with the KT Eri light curve confirms
that our formulation of free-free emission model light curve, Equation
(\ref{free-free_flux_v-band}), is reasonable.

\subsection{V339 Del 2013}
\label{v339_del}

% Fig.3
\begin{figure}
%\epsscale{0.7}
\epsscale{1.15}
%\plotone{v339del.f.ps}
\plotone{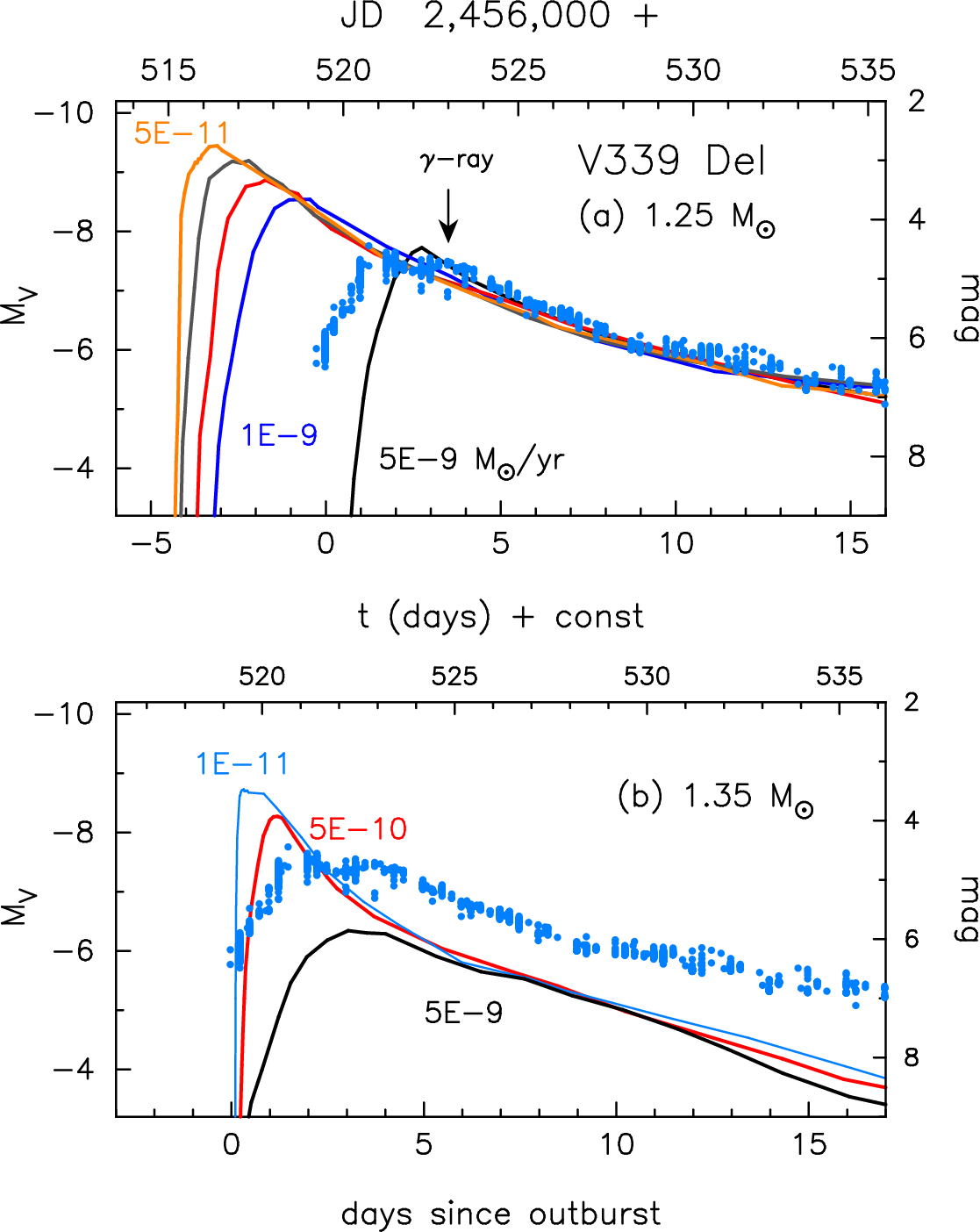}
\caption{Comparison between the V339 Del $V$ and $y$ light curves
and our model light curves.  The V339 Del $V/y$ data (cyan-blue dots) 
are all taken from Figure 1 of \citet{hac24km}.
Here, we assume $(m-M)_V=12.2$ for V339 Del.
(a) Fitting with the 1.25 $M_\sun$ WD models. 
The model light curve is shifted horizontally in the same way as 
in Figure \ref{kteri}.
%$t=0$ of each model is shifted horizontally 
%to fit with the steady-state part (four days after the outburst).
(b) Fitting with the 1.35 $M_\sun$ WD models. 
Three light curves are set at $t=0$. 
%The black dots show a 1 mag-down light curve of V339 Del.
}\label{v339del}
% source rn/m125.5e-9.C10.atL3/v339del.f.wip
\end{figure}

The classical nova V339 Del was discovered by Itagaki at 6.8 mag
on UT 2013 August 14.584 \citep[$=$JD 2,456,519.084,][]{nak13s}.
Immediately after the discovery, it was well observed in multiwavelength
bands, especially in photometry, because of its brightness 
(a naked-eye nova; $V\sim 4.4$ at peak).
This is the first nova in which $^7$Be lines are detected \citep{taj15sn}:  
The first definit observational evidence of a thermonuclear runaway. 
%on JD~2,456,585.5

GeV gamma-rays  were detected with the Fermi/LAT, first
on JD~2,456,523 at the epoch depicted by the downward arrow 
in Figure \ref{v339del}(a) \citep{ack14aa}, and up to JD~2,456,547.
The gamma-ray emission is closely related with a strong shock 
\citep[][for a recent review]{cho21ms}.
\citet{hac24km} analyzed multiwavelength light curve in detail 
including $V$, $y$, $K_s$ of \citet{mun15mm}, \citet{bur15a}, SMARTS
\citep{wal12bt}, and their own data, and X-ray light curves 
\citep[Swift: ][]{eva09},
and concluded that an optically thin dust-shell forms 
behind the shock. The epoch of dust formation, JD~2,456,553, is 
out of the range in Figure \ref{v339del}.  
Our nova evolution model does not include gamma-ray emission nor 
dust formation. 
However, the shock is formed far outside of the WD photopshere which 
does not affect our free-free emission light curve around its peak. 
Gamma-ray emission and shock formation will be discribed in Discussion 
(Section \ref{dis_shock}). 

\citet{hac24km} analyzed multiwavelength light curve of 
V339 Del and obtained $M_{\rm WD}=1.25 \pm 0.05~M_\sun$,  
$(m-M)_V =12.2 \pm 0.2$, and $d=2.1 \pm 0.2$ kpc for 
$E(B-V)=0.18$. 
Figure \ref{v339del}(a) shows $y$ and $V$ light curves 
assuming $(m-M)_V =12.2$ together with our 1.25 $M_\sun$ WD models. 
These model light curves are shifted in the horizontal direction to 
match the decay phase of each model.  
Among the five $\dot M_{\rm acc}$ models, the largest mass-accretion rate of
$\dot M_{\rm acc}=5 \times 10^{-9}~M_\sun$ yr$^{-1}$ is reasonably 
fitted with the V339 Del light curve. 
The slight difference in the epoch of the rising phase
indicates that the true mass-accretion rate is slightly smaller
than $5 \times 10^{-9}~M_\sun$ yr$^{-1}$, say, 
$\dot{M}_{\rm acc} \sim 3 \times 10^{-9}~M_\sun$ yr$^{-1}$.

Figure \ref{v339del}(b) shows the comparison between our 
1.35 $M_\sun$ WD models and the V339 Del light curve.
The cyan-blue data are the same as those in Figure \ref{v339del}(a). 
None of the 1.35 $M_\sun$ WD models can be fitted with the V339 Del
light curve. 
%If we shift down the cyan-blue data by 1 mag (black dots), 
The decline slope roughly agrees, but slightly slower than 
the theoretical lines. This indicates that the WD mass is 
less massive than 1.35 $M_\sun$. 

The good agreement with the 1.25 $M_\sun$ WD model is consistent with 
the estimates of $1.25 \pm 0.05 ~M_\sun$ of \citet{hac24km} obtained 
from fitting with the post-maximum phase of $V/y$ light curves. 
Their model light curves are based on the sequence of steady-state 
envelope solutions as introduced in Section \ref{optical_light_curve}
and Figure \ref{all}, but including the effects of 
an optically-thin shocked shell.  

%V339 Del shows a multiple peak followed by a smooth decline.

%For comparison we show our 1.35 $M_\sun$ models in panel (b). 
%The $\dot M_{\rm acc}=5 \times 10^{-9}~M_\sun$ yr$^{-1}$ model also 
%reproduce the light curve around the peak.

%All the models decay faster than V339 Del. 
%So the total nova equation timescale does not match to 1.35 $M_\sun$. 
%Thus we confirm that the WD mass 1.25 $M_\sun$ and the distance modulus 
%obtained by \citet{hac24km}. 

\subsection{V597 Pup 2007\#1} \label{sec_v597pup}

% Fig.4
\begin{figure}
%\epsscale{0.7}
\epsscale{1.15}
%\plotone{v597pup.f.ps}
\plotone{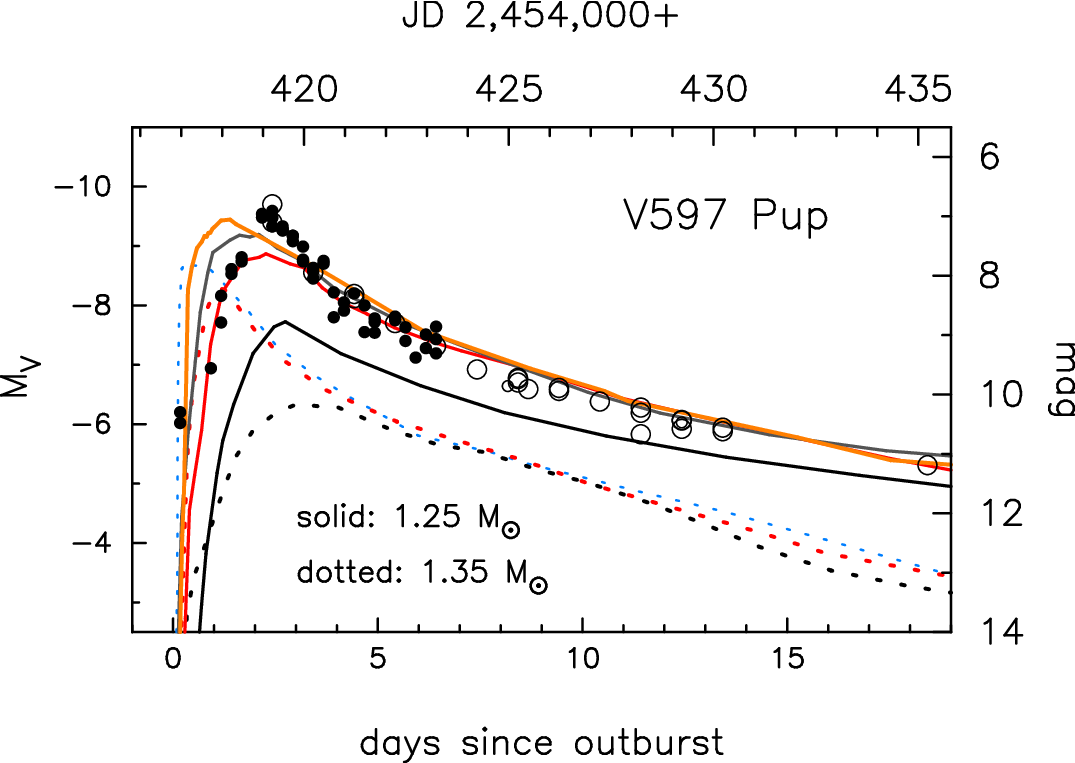}
\caption{Comparison between the $V$ light curve of the classical nova
V597 Pup and our model light curves for $(m-M)_V=17.0$. 
The optical data are the same as those in \citet{hac10k}. 
Filled circle: SMEI \citep{hou10bh}. Small open circle: AAVSO. 
Large open circle: VSOLJ. 
The solid lines are the model light curves for the 1.25 $M_\sun$ WD 
and the dotted lines are for the $1.35 ~M_\sun$ WD models. 
The mass accretion rate of each model is the same as that of 
the same color model in Figure \ref{all}, in which model M125ac1E-9 
is omitted to simplify the figure. 
}\label{v597pup}
% source: rn/m125.5e-9.C10.atL3/v597pup.f.wip
\end{figure}
 
V597 Pup was discovered by Pereira on 2007 November 14.23 UT
(JD 2454418.73) at mag about 7.0 \citep{per07}.
The optical data are obtained with SMEI before the 
optical maximum followed by VSOLJ and AAVSO data.
A bright supersoft X-ray phase was reported by \citet{nes08c}.
% Jan. 8.02 and 17.18 UT, 2008
The binary orbital period is 0.1112 day (2.6687 hr) by \citet{war09}

\citet{hac10k} analyzed the optical and supersoft X-ray light curves of 
V597 Pup and estimated the WD mass to be $1.2 \pm 0.05 ~M_\sun$. 
They also determined  $(m-M)_V = 17.0 \pm 0.2$ 
and the distance to V597 Pup, $d = 16 \pm 2$ kpc,
for $A_V = 3.1 E(B-V) = 3.1 \times 0.3 = 0.93$ \citep{nes08c}.

Figure \ref{v597pup} depicts the observed light curve of V597 Pup 
with $(m-M)_V = 17.0$. 
The 1.25 $M_\sun$ models show better fits than the 1.35 $M_\sun$ models 
(dotted lines) that are all fainter than the observed brightness. 
Among the $\dot M_{\rm acc}$ models of the 1.25 $M_\sun$ WD, the
$\dot M_{\rm acc}=5 \times 10^{-10}~M_\sun$ yr$^{-1}$ model (red line)
shows the closest fit.

The result is consistent with the WD mass estimate of 
$1.2 \pm 0.05 ~M_\sun$ and with the distance modulus in the $V$ band of
$(m-M)_V=17.0 \pm 0.2$ obtained by \citet{hac10k}.

\subsection{SMCN 2016-10a}\label{sec_smcn2016-10a}

% Fig.5
\begin{figure}
%\epsscale{0.7}
\epsscale{1.15}
%\plotone{smc2016.ps}
\plotone{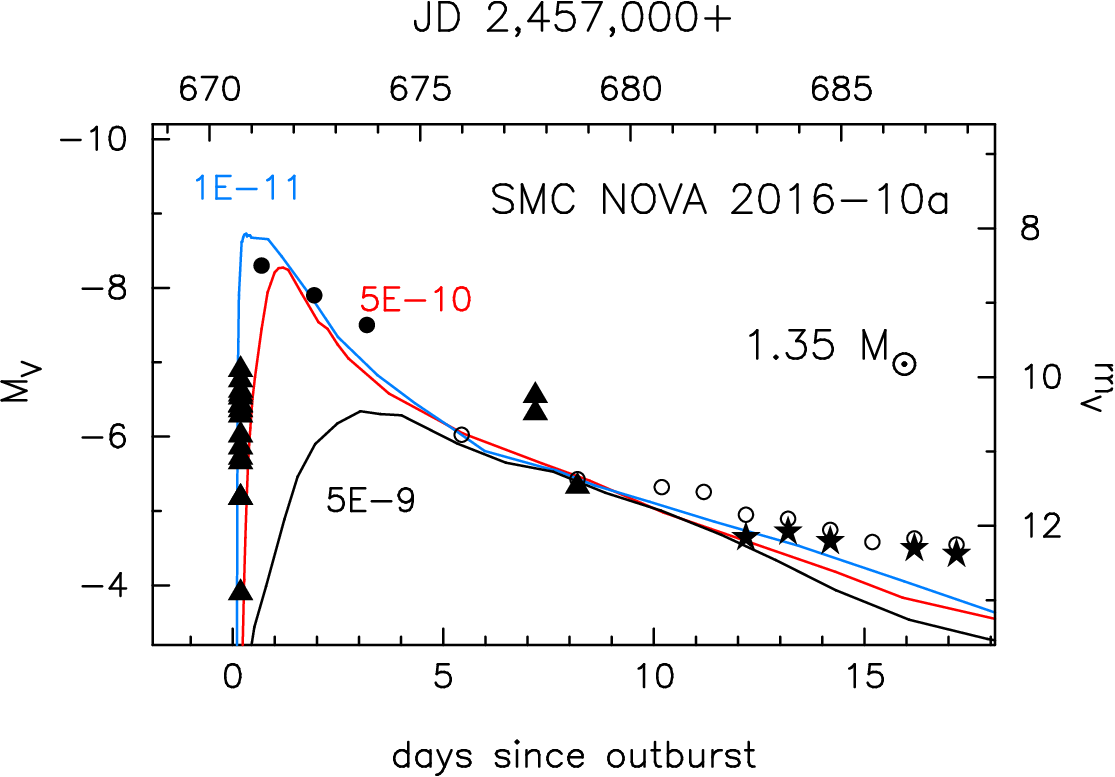}
\caption{Comparison between the $V$ light curve of the classical nova
SMC Nova 2016-10a and our model light curves of the 1.35 $M_\sun$ WD
models for $(m-M)_V=16.8$.  
The different symbols indicate different sources;  
Filled triangle: ATel No.9684 \citep{job16}, 
filled circle: ATel No.9631 \citep{lip16},
open circle: \citet{mun17}, 
and filled stars: SMARTS \citep{wal12bt}.
The reasonable fits by $(m-M)_V=16.8$ indicate that SMC Nova 2016-10a
is not a member of the SMC \citep[$(m-M)_V=19.0$,][]{ayd17},
but a foreground Galactic nova. 
}\label{smc2016}
% source: rn/m125.5e-9.C10.atL3/smc2016.wip
\end{figure}

The classical nova SMCN 2016-10a appeared 
in the direction of $(l,b)=(301^{\circ}.6362, -42^{\circ}.3037)$
toward the outskirt of the SMC. 
Thus, \citet{mun17} and \citet{ayd17} assumed that 
this object belongs to the SMC. 
On the other hand, \citet{hac18kb} 
obtained $(m-M)_V=16.8$ and $E(B-V)=0.08$ using the color-magnitude 
$(B-V)_0$-$M_V$ diagram and time-stretching method for nova light curves 
\citep[see Figure 36(c) of ][]{hac18kb}.
Then, the distance is $d=20.5$ kpc. 
They concluded that the classical nova SMCN 2016-10a
is a foreground Galactic object.  
No orbital periods have ever been reported.

Here, we adopt $(m-M)_V=16.8$ and plot our model light curves
as well as the SMCN 2016-10a observation in Figure \ref{smc2016}.  
The observational data are roughly consistent with the 1.35 $M_\sun$ WD
model with the mass accretion rate of
$\dot M_{\rm acc}=5 \times 10^{-10}~M_\sun$ yr$^{-1}$. 
This confirms that the distance $d=20.5$ kpc is reasonable
and SMCN 2016-10a is not a member of the SMC, but a foreground 
Galactic object. 

\citet{hac18kb} also analyzed light curves of $V$, $I_C$,
and supersoft X-ray light curves and suggested that 
the WD mass is about 1.29 $M_\sun$. 

The light curve of SMCN 2016-10a is broadly consistent with 
the 1.35 $M_\sun$ with the mass-accretion rate of
$\dot M_{\rm acc}=1 \times 10^{-11}~M_\sun$ yr$^{-1}$ 
or $\dot M_{\rm acc}=5 \times 10^{-10}~M_\sun$ yr$^{-1}$.
It should be noted that our model in Figure \ref{smc2016}
is calculated with the solar heavy elements composition of 
$Z=0.02$ besides the extra carbon enrichment of $X_C=0.1$.  
Considering its location, 
far from the Galactic plane at $z=-13.75$ kpc,  
this nova is possibly a population II star ($Z \leq 0.004$). 
If we take into account the tendency that a lower metallicity $Z$ nova shows 
a slower decline \citep{kat13hh},
the WD mass of SMCN 2016-10a could be more massive than 1.35 $M_\sun$.

%%%%%%%%%%%%%%%%%%%%%%%%%%%%%

\section{Discussion}\label{sec_discussion}

% Fig.6
\begin{figure}
%\epsscale{0.7}
\epsscale{1.15}
\plotone{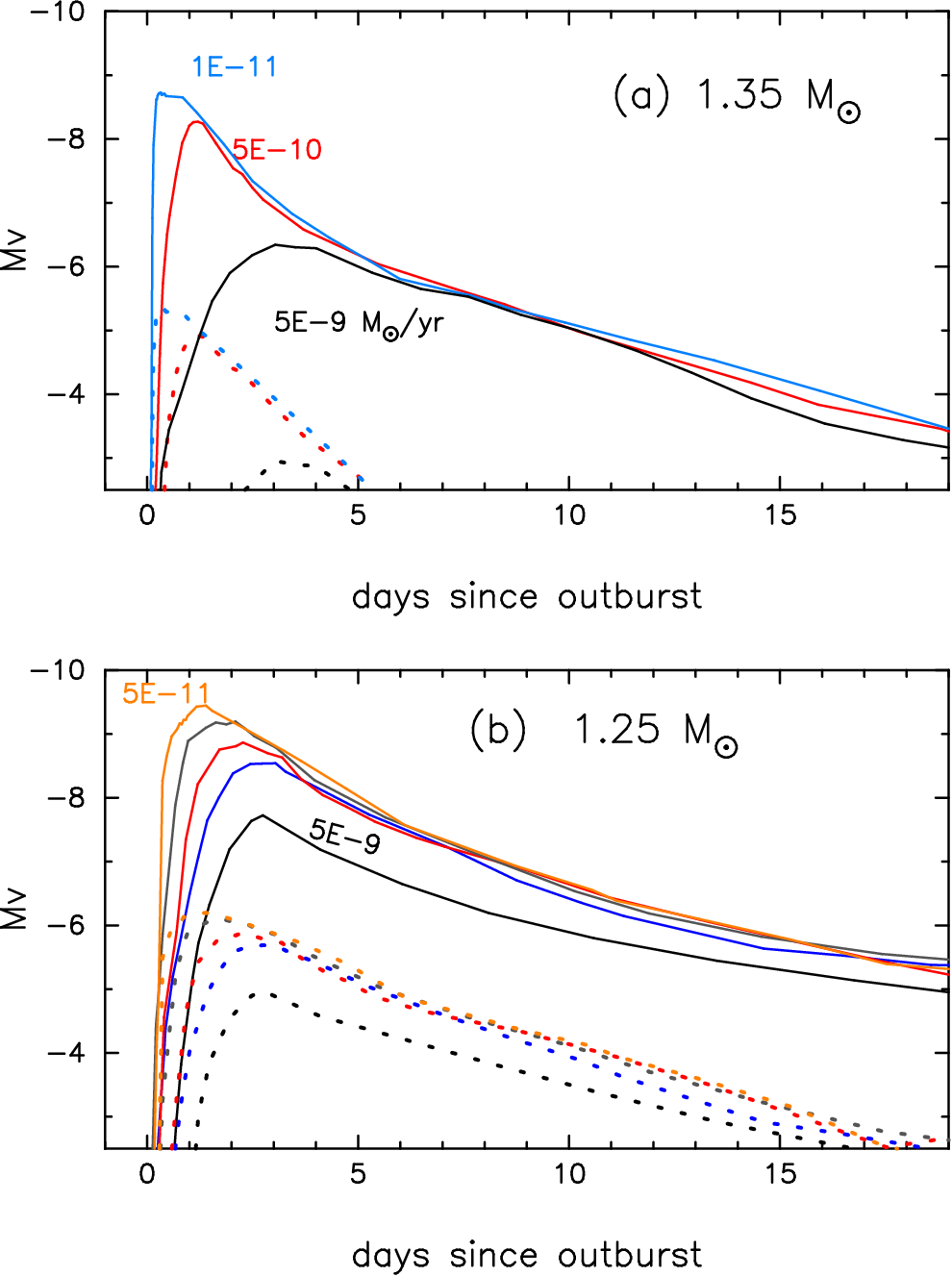}
\caption{Comparison of $V$ light curves between the free-free emission
from the nova winds and the blackbody emission of the WD photosphere
for each model.  The models denoted by the thick lines are the same as
in Figure \ref{all}.  The dotted lines denote the $V$ light curves
for blackbody emission of the WD photosphere corresponding to the model
with the same color. 
}\label{BB}
% source: rn/m125.5e-9.C10.atL3/all.bb.wip
\end{figure}

\subsection{Emission from the WD photosphere}\label{dis_bb}

In the present work, we have neglected the emission from the WD photosphere 
in calculating our theoretical $V$ light curves. 
 
\citet{hac15k} analyzed 11 slow/fast novae and clarified that 
the free-free emission from the optically thin ejecta just outside 
of the WD photophsere always dominates the WD photospheric emission
(see their Figures 28 and 29), where they assumed blackbody approximation
for the WD photospheric emission. 
\citet{hac25kv1674her2} clearly showed that the photospheric blackbody
emission is rather small in the very fast nova V1674 Her (see their Figure 6).
%This difference between the free-free and blackbody emission is large 
In fast novae, the flux of free-free emission substantially exceeds
the flux of photospheric blackbody emission
becasue their wind mass loss rates are large. 
In V1674 Her, V1500 Cyg, GK Per, and V603 Aql, 
the luminosity of the free-free emission is one order of magnitude 
larger ($\sim 2.5$ mag brighter) than that of the photospheric
blackbody emission \citep{hac15k, hac25kv1674her2}.  

%
% in PW Vul, V705 Cas, GQ Mus, RR Pic, V5558 Sgr, HR Del, V723 Cas,
% GK Per (0.77 Mo), V603 Aql (1.2 Mo), V1500 Cyg, DQ Her 
%

Figure \ref{BB} shows the light curves of the blackbody emission for 
all of our theoretical models.
% (dotted lines) in Figure \ref{all}.
They are much fainter than the free-free luminosities. 
Thus, our assumption that we do not include the WD photospheric
emission is appropreate.

\subsection{Shock Formation and Gamma-Ray Emission} \label{dis_shock}

GeV-gamma rays and hard X-rays have been often observed in nova outbursts,
which are considered to originate from a strong shock 
between shells ejected with different velocities
\citep[e.g.,][]{muk01i, abd10, ack14aa, che16, cho14ly, fin18cm,
gor21ap, mar18dj, met15fv, nel12a, sok22ll}.
If the inner shell (later ejected) has a larger velocity than that of
the outer shell (earlier ejected), the inner one can catch up with the
outer one and forms a shock wave \citep[e.g.,][]{muk19s, ayd20ci, ayd20sc}.
The assumption of multiple shell ejection is the key of this idea.

From the theoretical point of view, however, 
many numerical nova outburst calculations have been presented
from the early thermonuclear runaway to the extended phase of
nova outbursts \citep[e.g.,][]{pri92k, pri95k, epels07, sta09ih, den13hb,
chen19, kat22sha, kat22shapjl}, which show that
mass ejection is continuous, no multiple mass ejection occurs, therefore, 
no shock arises at the thermonuclear runaway.

\citet{hac22shock} presented a shock formation mechanism  
based on \citet{kat22sha}'s
fully self-consistent nova explosion model.
A strong reverse shock inevitably arises far outside of 
the WD photosphere after the optical peak because 
the later ejected matter has a larger expansion velocity 
that catches up the former ejected matter  
\citep[see, e.g., Figure 1 of ][]{hac22shock}.
\citet{hac23k} clarified that the shock 
devides expanding ejecta into three regions with different velocities, 
which naturally explains multiple velocity emission/absorption line systems
\citep{mcl42, fri87}.
This shock formation mechanism reasonably explains gamma-ray emission
and hard X-ray detection/nondetection in classical novae \citep[e.g.,
YZ Ret, V339 Del, and V392 Per in][respectively]{hac23k, hac24km,
hac25kv392per}.

Among the four novae studied in this paper, V339 Del is a gamma-ray nova. 
\citet{hac24km} presented multiwavelength light curve model. 
The overall $V$ and $y$ light curves are explained by free-free emission 
from the ejcta just outside of the WD photophsere. 
The shocked shell is formed far outside the WD photopshere and its luminosity 
is one order of magnitude smaller than the free-free emission luminosity 
in the first 20 days
%, i.e., $V$ magnitude is about 2.5 mag fainter 
\citep[see Fig. 8(a) of ][]{hac24km}.
They also explained the gamma-ray light curve
from the optically thin shoked matter.

As the shocked shell is optically thin and its brightness is much smaller than 
the free-free emission, i.e., Equation (\ref{free-free_flux_v-band}), 
the light curve around the peak is not 
affectd by the presence/non-presence of the shocked shell. 
We thus assumed that our optical light curves can be applied to
various novae independently of gamma-ray detection.  
We reasonably reproduced the light curves of KT Eri, V339 Del, V597 Pup, 
and SMC N 2016-10a only by our free-free emission models.
This strongly supports that free-free emission dominates
the nova luminosities if the shocked shell is optically thin.

\subsection{Spherical Symmetry}

We have assumed spherical symmetry in our nova outburst calculation. 
It is well known that some novae show asphericity of the ejecta 
\citep[see, e.g., ][for old nova shells]{dow00}. 
%%%\citep[for RS Oph][]{lic24} ( more ref). 
As these images were taken much later than the optical peak, 
they are not the direct evidence of aspherical expansion
at the epoch of optical peak. 

It has been discussed that the sphericity (or elongation) of a nova shell
depends on its (nova) speed class.  The faster the nova speed class is,
the more spherical the shell is \citep[e.g.,][]{dow00}.
KT Eri \citep[$t_2=6.6$ days, ][]{hou10bh}, 
V339 Del \citep[$t_2=10$ days, ][]{cho14},
V597 Pup \citep[$t_2=2.5$ days, ][]{naik09}, and
SMC N 2016-10a \citep[$t_2=4.0$ days, ][]{aydi18}
belong to the very fast novae
class \citep[$t_2\le 10$ days, ][]{pay57} and
its shell is probably close to spherical. Therefore, our assumption
of spherical symmetry does not so largely depart from the true shell.
This can be understood by the fact that a faster speed class of novae have
larger expansion velocities and therefore their ejecta have less interaction
times with the orbital motion of a companion star.

Before the WD photosphere reaches the companion star,
its photospheric radius is as small as $\lesssim 0.1 ~R_\sun$ and
therefore is observed as an X-ray flash \citep{kon22wa, kat22shapjl}. 
Such a case was seen in YZ Ret 2020, in which an X-ray flash is first 
detected.  The X-ray luminosity during the flash phase, calculated from the 
observed blackbody temerature assuming spherical photosphere, 
is very consistent with the Eddington luminosity of a 1.35 $M_\sun$ WD
\citep{kon22wa, kat22shapjl}.  This consistency supports that the WD 
photosphere is almost spherical and its shape does not much deviate
from a sphere.  In other words, this X-ray flash observation
is the first direct indication of spherically symmetric expansion 
of the photopsere before the optical peak. 

After the WD photosphere engulfs the companion star, free-free emission
from the nova wind dominates the $V$ luminosity.  For one of the fastest
class of novae, V1674 Her 2021 \citep[$t_2=0.9$ days, ][]{hab24},
\citet{hac25kv1674her2} showed that their free-free emission model 
$V$ light curve calculated with Equation (\ref{free-free_flux_v-band})
for a 1.35 $M_\sun$ WD with $\dot{M}_{\rm acc}=1\times 10^{-11} ~M_\sun$
yr$^{-1}$ almost perfectly reproduced the observed $g$ band light curve
in the rising phase from $g=$15 to 7 mag
\citep[see Figure 6 of ][]{hac25kv1674her2}.
This also confirms that the WD ejecta is almost spherically symmetric.
In this case, the rising time from $g=$15 to 7 mag is only 0.2 days,
which is close to the orbital period of 0.153 days \citep{pat22}.
There seems to be almost no time to interact with the orbital motion.

\subsection{Comparison with steady-state models}

The dotted black line in Figure \ref{all}(a) shows the light curve constructed from
a time-sequence of steady-state envelope solutions on a 1.35  $M_\sun$ WD 
with uniform chemical composition \citep{hac06kb,hac25kv392per}.
This dotted line agrees well with the post-maximum phase
of all the three fully self-consistent nova models. 
This indicates that in the post-maximum phase thermonuclear runaway almost 
settled down and the nuclear energy generation rate becomes almost 
constant with time \citep[see, e.g., Figure 2 in ][]{kat25v1674her}. 
%The WD envelope reaches the thermal equilibrium 

Figure \ref{kteri}(a) shows the steady-state sequence model 
of the $1.25~M_\odot$ with the chemical composition of $X=0.6$,
$Y=0.25 $, $Z_{\rm CNO}=0.13$, and $Z=0.02$ and the radius of 
the envelope to be $\log R/R_\sun=-2.29$.
%, taken from model M125ac5E-9.  
Our model light curve follows well the steady-state sequence model 
in the post-maximum phase. 

To summarize, soon after the optical peak, a nova explosion evolution
can be approximated by a sequence of steady-state solutions with 
decreasing envelope mass 
as already reported in a fully self-consistent nova explosion 
model on a 1.0 $M_\sun$ WD \citep{kat22sha}. 

\citet{kat83} proposed an idea that 
nova explosions can be followed by a sequence of steady-state 
optically-thick wind solutions with the decreasing envelope mass 
because a nova envelope would settle down to a steady state 
shortly after a thermonuclear runaway reaches maximum. 
This optically thick nova wind theory was formulated by \citet{kat94h}.
% A model sequence with decreasing envelope mass is regarded 
% as the sequence of evolution models with losing mass by 
% the optical-thick winds. 
%
Based on the steady-state envelope sequences, \citet{hac06kb} calculated 
a number of free-free emission model nova light curves. 
%found remarkable similarity in nova light curves; 
Many novae have been analyzed with these free-free emission model
nova light curves and their WD masses have been determined. 
The present work on the 1.25 $M_\sun$ and 1.35 $M_\sun$ WDs gives
an additional theoretical support to these nova light curve analyses.

%All the light curves can be reduced to a standard curve
%independent of the WD mass, chemical composition of the envelope, 
%and frequency (IR and optical bands) if we choose an appropriate parameter 
%(stretching factor in Appendix \ref{sec_timestretching}). 
%They call it the ``universal decline law'' of nova light curves.  

\section{Conclusion}\label{sec_conclusion}

Our results are summarized as follows: \\
\begin{enumerate}
\item We present theoretical light curves of classical novae for 
1.25 $M_\sun$ and 1.35 $M_\sun$ WDs. Our numerical code is a
Henyey-type evolution code combined with the optically thick winds. 
The optical $V$ light curve is calculated for free-free emission.  
Our $V$ light curves predict nova light curves,  
from a rising phase toward optical maximum and the early decay phase. 
\item We have compared our light curves with those of the classical novae,
KT Eri, V339 Del, V597 Pup, and SMC Nova 2016-10a
whose WD masses and distance moduli in the $V$ band are well constrained. 
Our models show excellent agreements with these novae. 
%
%\item The 1.25 $M_\sun$ WD models well reproduce KT Eri, V339 Del,
%and V597 Pup while the 1.35 $M_\sun$ model does SMCN 2016-10a. 
 The WD masses of these novae are consistent with the results
 in Table \ref{table_novae}, 
which are independently obtained from the entire decay phase of novae
including the X-ray light curve fittings of their SSS phases. 
\item These good agreements with the $V$ light curves support that the 
rising phase toward optical maximum is well approximated by
our free-free emission model light curves. 
%
% \item The well fitting around the peak confirmed that our calculation method 
% and distance modulus $(m-M)_V$ is reasonable. 
%
\item We confirm that the time-sequence of steady-state envelope solutions
with the decreasing envelope mass is a good approximation to nova outburst 
explosions on the 1.25 $M_\sun$ and 1.35 $M_\sun$ WDs
in the post-maximum phase. 
\end{enumerate}

\begin{acknowledgments}
 We thank the American Association of Variable Star Observers
(AAVSO) for the archival data of KT Eri, V597 Pup
and the Variable Star Observers League of Japan (VSOLJ) for V597 Pup. 
 We are also grateful to the anonymous referee for useful comments
regarding how to improve the manuscript.
\end{acknowledgments}

%\vspace{5mm}
%\facilities{Swift(XRT), AAVSO}

%%%%  Appendix A

%\appendix

% todayreference


\begin{thebibliography}{}

\bibitem[A. A. Abdo et al. (2010)]{abd10}
Abdo, A. A., Ackermann, M., Ajello, M., et al. 2010, Science,  329, 817,
\doi{10.1126/science.1192537}
%-821.  Gamma-Ray Emission Concurrent with the Nova
%  in the Symbiotic Binary V407 Cygni

\bibitem[M. Ackermann et al. (2014)]{ack14aa}
Ackermann, M., Ajello, M., Albert, A., et al. 2014, Science, 345, 554,
\doi{10.1126/science.1253947}
%-558.  Fermi establishes classical novae as a distinct class
%  of gamma-ray sources

%\bibitem[E. Aydi et al. (2023)]{ayd23}
%Aydi, E., Chomiuk, L., Miko\l ajewska, J., et al. 
%%%Brink, J., Metzger, B.D. et al. 
%2023, \mnras, 524, 1964, \doi{10.1093/mnras/stad1914}
% Chatching a nova X-ray/UV flash in the visible? Early spectroscopy of
% very slow nova velorum 2022 (Gaia22alz)

\bibitem[E. Aydi et al. (2017)]{ayd17}
Aydi, E.,  Page, K. L., Kuin, N. P. M., et al. 2017, \mnras, 474, 2679, 
\doi{10.1093/mnras/stx2678}
%(arXiv: 1710.03716)
%  Multiwavelength observations of nova SMCN 2016-10a
% --- one of the brightest novae ever observed

\bibitem[E. Aydi et al. (2018)]{aydi18}
Aydi, E.,  Page, K. L., Kuin, N. P. M., et al. 2018, \mnras, 474, 2679,
\doi{10.1093/mnras/stx2678}
%%(arXiv: 1710.03716)
%  Multiwavelength observations of nova SMCN 2016-10a
% --- one of the brightest novae ever observed
% Multiwavelength observations of nova SMCN 2016-10a 
% - one of the brightest novae ever observed
%Aydi, E. ; Page, K. L.; Kuin, N. P. M. ; Darnley, M. J.; Walter, F. M.;
%Mroz, P.; Buckley, D. A. H.; Mohamed, S.; Whitelock, P.; Woudt, P.;
%Williams, S. C.; Orio, M.; Williams, R. E.; Beardmore, A. P.;
%Osborne, J. P.; Kniazev, A.; Ribeiro, V. A. R. M.; Udalski, A.;
%Strader, J.; Chomiuk, L.
% Monthly Notices of the Royal Astronomical Society, Volume 474,
% Issue 2, p.2679-2705 
% Pub Date: February 2018 DOI: 10.1093/mnras/stx2678 

\bibitem[E. Aydi et al. (2020a)]{ayd20ci}
Aydi, E., Chomiuk, L., Izzo, L., et al. 2020a, \apj, 905, 62,
\doi{10.3847/1538-4357/abc3bb}
% Early Spectral Evolution of Classical Novae:
% Consistent Evidence for Multiple Distinct Outflows
% The Astrophysical Journal, Volume 905, Issue 1, id.62, 33 pp.
% Pub Date: December 2020
% 10.3847/1538-4357/abc3bb

\bibitem[E. Aydi et al. (2020b)]{ayd20sc}
Aydi, E., Sokolovsky, K. V., Chomiuk, L., et al. 2020b,
Nature Astronomy, 4, 776, \doi{10.1038/s41550-020-1070-y}
% Direct evidence for shock-powered optical emission in a nova
% Nature Astronomy, Volume 4, p. 776-780
% Pub Date: April 2020
% 10.1038/s41550-020-1070-y

\bibitem[C. A. L. Bailer-Jones et al. (2021)]{bai21gaia}
Bailer-Jones, C.A.L., Rybizki, J., Fouesneau, M., Demleitner, M., \& Andrae,R. 
2021, \aj, 161, 147, \doi{10.3847/1538-3881/abd806}
% Estimating distances from parallaxes, V. Geometric 
%and photogeometric distances to 1.47 billion Stars in Gaia Early 
%Data Release 3

%\bibitem[Y. Bhargava et al. (2024)]{bha24}
%Bhargava, Y., Dewangan, G. C., Anupama, G.,C., et al.
%2024, \mnras, 528, 28, \doi{10.1093/mnras/stad3870}
% Soft X-ray and FUV observation of nova her 2021 (V1674 Her) with Astrosat

%\bibitem[H. Bondi (1952)]{bon52}
%Bondi, H. 1952, \mnras, 112, 195, \doi{10.1093/mnras/112.2.195}
% On spherically symmetrical accretion

\bibitem[D. A. H. Buckley \& Tuohy (1989)]{buc89}
Buckley, D. A. H., \& Tuohy, I.R. 1989, \apj, 344, 376, \doi{10.1086/167806}
% A Spectroscopic, photometric, & X-ray study of the DQ Her system 1H0542-407

\bibitem[M.A. Burlak et al. (2015)]{bur15a}
Burlak, M. A., Esipov, V. F., Komissarova, G. V., et al. 2015,
Baltic Astronomy, 24, 109, \doi{10.1515/astro-2017-0209}
%-352.   UBVJHKLM photometry and low-resolution spectroscopy of
% Nova Delphini 2013 (V339 Del)

\bibitem[H.-L. Chen et al. (2019)]{chen19}
Chen, H.-L., Woods, T. E., Yungelson, L. R., et al.
2019, \mnras, 490, 1678, \doi{10.1093/mnras/stz2644}
%% arXiv:1807.02241
% Comprehensive models of novae at metallicity Z = 0.02 and Z = 10^-4
% Chen, Hai-Liang ; Woods, T. E. ; Yungelson, L. R. ; Piersanti, Luciano;
%Gilfanov, M. ; Han, Zhanwen
%Monthly Notices of the Royal Astronomical Society, Volume 490,
% Issue 2, p.1678-1692
% Pub Date: December 2019 DOI: 10.1093/mnras/stz2644

\bibitem[C. C. Cheung et al. (2016)]{che16}
Cheung, C. C., Jean, P., Shore, S. N., et al., \apj, 826, 142
\doi{10.3847/0004-637X/826/2/142}
%, 12 pp.   Fermi-LAT Gamma-Ray Detections of Classical Novae
% V1369 Centauri 2013 and V5668 Sagittarii 2015

%\bibitem[M. della Valle \& M. Livio (1995)]{del95}
%della Valle, M., \& Livio, M. 1995, \apj, 452, 704, \doi{10.1086/176342}
%-709. The Calibration of Novae as Distance Indicators

\bibitem[D. Chochol et al. (2014)]{cho14}
Chochol, D., Shugarov, S., Pribulla, T., Volkov, I. 2014, CoSka, 43, 330
% Photometry and spectroscopy of the classical nova V339 Del
% (Nova Del 2013) in the first month after outburst
% no doi

\bibitem[L. Chomiuk et al. (2014)]{cho14ly}
Chomiuk, L., Linford, J. D., Yang, J., et al. 2014, \nat, 514, 339,
\doi{10.1038/nature13773}
%Nature, Volume 514, Issue 7522, pp. 339-342 (2014).
% Binary orbits as the driver of gamma-ray emission and mass ejection
% in classical novae

\bibitem[L. Chomiuk et al. (2021)]{cho21ms}
Chomiuk, L., Metzger, B. D., \& Shen, K. J. 2021,
Annual Review of Astronomy and Astrophysics, 59, 48,
\doi{10.1146/annurev-astro-112420-114502}
%New Insights into Classical Novae
%(arXiv:2011.08751)

\bibitem[P.A. Denissenkov et al. (2013)]{den13hb}
Denissenkov, P. A., Herwig, F., Bildsten, L., \& Paxton, B.
2013, \apj, 762, 8 \doi{10.1088/0004-637X/762/1/8}
% The Astrophysical Journal, Volume 762, Issue 1, article id. 8, 10 pp. (2013).
% MESA Models of Classical Nova Outbursts:
% The Multicycle Evolution and Effects of Convective Boundary Mixing

\bibitem[R. A. Downes \& H. W. Duerbeck (2000)]{dow00}
Downes, R. A., \& Duerbeck, H. W. 2000, \aj, 120, 2007, \doi{10.1086/301551}
%-2037. Optical Imaging of Nova Shells and the Maximum Magnitude-Rate
% of Decline Relationship

%\bibitem[J. J. Drake et al. (2021)]{dra21}
%Drake, J. J., Ness, J.-U., Page, K. L., et al. 2021, \apjl, 922, L42,
%\doi{10.3847/2041-8213/ac34fd}
% spin-down and ultrafast outflows of the highly pulsed SSS
% of nova Her 2021  (V1674 Her)

\bibitem[D. Ennis et al. (1977)]{enn77}
Ennis, D., Becklin, E. E., Beckwith, S., et al.
%%% Elias, J., Gatley, I., Matthews, K., Neugebauer, G., \& Willner, S. P.
1977, \apj, 214, 478, \doi{10.1086/155273}
%%-487. Infrared observations of Nova Cygni 1975

\bibitem[N. Epelstain et al. (2007)]{epels07}
Epelstain, N., Yaron, O., Kovetz, A., \& Prialnik, D. 2007, \mnras, 374, 1449,
\doi{10.1111/j.1365-2966.2006.11254.x}
%   A thousand and one nova outbursts

\bibitem[P. A. Evans et al. (2009)]{eva09}
Evans, P. A., Beardmore, A. P., Page, K. L., et al.  2009, \mnras, 397, 1177,
\doi{10.1111/j.1365-2966.2009.14913.x}
%-1201. (arXiv: 0812.3662)
%  Methods and results of an automatic analysis
% of a complete sample of Swift-XRT observations of GRBs

\bibitem[T. Finzell et al. (2018)]{fin18cm}
Finzell, T., Chomiuk, L., Metzger, B. D., et al. 2018, \apj, 852, 108,
\doi{10.3847/1538-4357/aaa12a}
% A Detailed Observational Analysis of V1324 Sco,
% the Most Gamma-Ray-luminous Classical Nova to Date

\bibitem[M. Friedjung (1966)]{fri66}
Friedjung, M. 1966, \mnras, 132, 317, \doi{10.1093/mnras/132.2.317}
% the physics of the nova phenomenon III  (radiation pressure)

\bibitem[M. Friedjung (1987)]{fri87}
Friedjung, M. 1987, \aap, 180, 155
%The formation of principal system of novae
% no doi number

\bibitem[J. S. Gallagher \& E. P. Ney (1976)]{gal76}
Gallagher, J. S., \& Ney, E. P. 1976, \apj, 204, L35, \doi{10.1086/182049}
%%-L39. The early infrared development of Nova Cygni 1975.

\bibitem[R. D. Gehrz et al. (1998)]{geh98tw}
Gehrz, R. D., Truran, J. W., Williams, R. E., \& Starrfield, S. 1998,
\pasp, 110, 3,  \doi{10.1086/316107}
%%-26. Nucleosynthesis in Classical Novae and Its Contribution
%% to the Interstellar Medium

\bibitem[A. C. Gordon et al. (2021)]{gor21ap}
Gordon, A. C., Aydi, E., Page, K. L., et al. 2021, \apj, 910, 134,
\doi{10.3847/1538-4357/abe547}
% Li, Kwan-Lok ; Chomiuk, L. ; Sokolovsky, K. V. ; Mukai, K. ; Seitz, J.
%The Astrophysical Journal, Volume 910, Issue 2, id.134, 19 pp.
% Pub Date: April 2021 DOI: 10.3847/1538-4357/abe547
%Surveying the X-Ray Behavior of Novae as They Emit Gamma-Rays

\bibitem[G. M. Green et al. (2019)]{gre19}
Green, G. M., Schlafly, E. F., Zucker, C., et al. 2019, \apj, 887, 93
\doi{10.3847/1538-4357/ab5362}
%%%arXiv:1801.03555
%% A Three-Dimensional Map of Milky-Way Dust

\bibitem[G. R. Habtie et al. (2024)]{hab24}
Habtie, G. R. Das, R., Pandey, R., Ashok, N.M., \& Dubovsky, P.A. 2024,
\mnras, 527, 1405, \doi{10.1093/mnras/stad3295}
% CN V1674 Her, element abundance

\bibitem[I. Hachisu \& M. Kato (2006)]{hac06kb}
Hachisu, I., \& Kato, M. 2006, \apjs, 167, 59
\doi{10.1086/508063}
%-80.  A Universal Decline Law of Classical Novae

\bibitem[I. Hachisu \& M. Kato (2010)]{hac10k}
Hachisu, I., \& Kato, M. 2010, \apj, 709, 680,
\doi{10.1088/0004-637X/709/2/680}
%-714. A Prediction Formula of Supersoft X-ray Phase of Classical Novae

\bibitem[I. Hachisu \& M. Kato (2015)]{hac15k}
Hachisu, I., \& Kato, M. 2015, \apj, 798, 76, \doi{10.1088/0004-637X/798/2/76}
% (1-29) A Light Curve Analysis of Classical Novae: Free-free Emission vs.
% Photospheric Emission

\bibitem[I. Hachisu \& M. Kato (2016)]{hac16k}
Hachisu, I., \& Kato, M. 2016, \apj, 816, 26, \doi{10.3847/0004-637X/816/1/26}
%%in press (arXiv:1511.06819), pp.69
% Light Curve Analysis of Neon Novae

\bibitem[I. Hachisu \& M. Kato (2018a)]{hac18k}
Hachisu, I., \& Kato, M. 2018a, \apj, 858, 108, \doi{10.3847/1538-4357/aabee0}
%16 pp. A Light-curve Analysis of Gamma-Ray Nova V959 Mon:
% Distance and White Dwarf Mass

\bibitem[I. Hachisu \& M. Kato (2018b)]{hac18kb}
Hachisu, I., \& Kato, M. 2018b, \apjs, 237, 4, \doi{10.3847/1538-4365/aac833}
%51 pp. A Light Curve Analysis of Recurrent and Very Fast Novae
% in our Galaxy, Magellanic Clouds, and M31

\bibitem[I. Hachisu \& M. Kato (2019a)]{hac19k}
Hachisu, I., \& Kato, M. 2019a, \apjs, 241, 4, \doi{10.3847/1538-4365/ab0202}
%68 pp. The UBV Color Evolution of Classical Novae. III.
% Time-Stretched Color-Magnitude Diagram of Recent Novae in Outburst
% (Paper III),

\bibitem[I. Hachisu \& M. Kato (2019b)]{hac19kb}
Hachisu, I., \& Kato, M. 2019b, \apjs, 242, 18, \doi{10.3847/1538-4365/ab1b43}
%138 pp. A Light Curve Analysis of 32 Recent Galactic Novae ---
% Distances and White Dwarf Masses

\bibitem[I. Hachisu \& M. Kato (2021)]{hac21k}
Hachisu, I., \& Kato, M. 2021, \apjs, 253, 27, \doi{10.3847/1538-4365/abd31e}
%%%in press (arXiv:2012.06100)
%  The $UBV$ Color Evolution of Classical Novae. IV.
%  Time-Stretched $(U-B)_0$-$(M_B-2.5\log f_{\rm s})$ and
%  $(V-I)_0$-$(M_I-2.5\log f_{\rm s})$ Color-Magnitude Diagrams
%  of Novae in Outburst

\bibitem[I. Hachisu \& M. Kato (2022)]{hac22shock}
Hachisu, I., \& Kato, M. 2022, \apj, 939, 1 \doi{10.3847/1538-4357/ac9475}
% nova full cycle model naturally produces a strong shock.
% it explains hard X-ray emission, principal, diffuse enhanced systems

\bibitem[I. Hachisu \& M. Kato (2023)]{hac23k}
Hachisu, I., \& Kato, M. 2023, \apj, 953, 78, \doi{10.3847/1538-4357/acdfd3}
%%%in press, arXiv:2306.09218
% A multiwavelength light curve analysis of the classical nova YZ Ret:
% An extension of the universal decline law to the nebular phase

\bibitem[I. Hachisu \& M. Kato (2025a)]{hac25kv392per}
Hachisu, I., \& Kato, M. 2025a, \apj, 984, 136, \doi{10.3847/1538-4357/adc38a}
% in press, arXiv:2503.13384
% V392 Per

\bibitem[I. Hachisu \& M. Kato (2025b)]{hac25kv1674her2}
Hachisu, I., \& Kato, M. 2025b, \apj, 989, 153,
\doi{10.3847/1538-4357/adef0b}
% Optical Detection of the X-Ray Flash in the Very Fast Nova V1674 Her:
% Optical Contribution of the Irradiated Accretion Disk
% DOI: 10.3847/1538-4357/adef0b

%\bibitem[I. Hachisu \& M. Kato (2025d)]{hac25kv1674her3}
%Hachisu, I., \& Kato, M. 2025d, \apj, 989, 153, \doi{10.3847/1538-4357/adef0b}
% Formation of an optically thick shocked shell in the very fast nova
% V1674 Herculis: the origin of superbrightness

\bibitem[I. Hachisu et al. (2024)]{hac24km}
Hachisu, I., Kato, M., \& Matsumoto, K. 2024, \apj, 965, 49,
\doi{10.3847/1538-4357/ad2a45}
%-19pp.  A multiwavelength light-curve model of the classical nova V339 Del:
% the coexistence of dust dip and supersoft X-rays

\bibitem[I. Hachisu et al. (2025)]{hac25kw}
Hachisu, I., Kato, M., \& Walter, F. M. 2025, \apj, 980, 142,
\doi{10.3847/1538-4357/adae08}
% arXiv:2412.00250
% A multiwavelength light curve analysis of the classical nova KT Eri: 
% Optical contribution from a large irradiated accretion disk

%\bibitem[I. Hachisu et al. (2016)]{hac16sk}
%Hachisu, I., Saio, H., \& Kato, M. 2016, \apj, 824, 22,
%\doi{10.3847/0004-637X/824/1/22}
% shortest recurrence period of forced novae, Idan's model

\bibitem[I. Hachisu et al. (2020)]{hac20skhs}
Hachisu, I., Saio, H., Kato, M., Henze, M., \& Shafter, A. W. 2020,
\apj, 902, 91, \doi{10.3847/1538-4357/abb5fa}
%%% (arXiv:2009.02937)
% A Theory for the Maximum Magnitude versus Rate of Decline
% (MMRD) Relation of Classical Novae

\bibitem[M. Henze et al. (2018)]{hen18}
Henze, M., Darnley, M., Williams, S. C. et al. 2018, \apj, 857, 68,
\doi{10.3847/1538-4357/aab6a6}
% Breaking the habit -- the peculiar 2016 eruption of the unique
%recurrent nova M31N 2008-12a

\bibitem[M. Henze et al. (2015)]{hen15}
Henze, M., Ness, J.-U., Darnley, M., et al. 2015, \aap, 580, 46
% (arXiv1504.06237)
% Henze, M.; Ness, J.-U.; Darnley, M. J.; Bode, M. F.; Williams,
% S. C.; Shafter,
% A. W., Sala, G., Kato, M.; Hachisu, I., and Hernanz, M.
% A remarkable recurrent nova in M31 :The predicted 2014 outburst in
% X-rays with Swift No2 paper

\bibitem[M. Henze et al. (2011)]{hen11} Henze, M., Pietsch, W., Haberl, F.,
Hernanz, M., Sala, G., Hatzidimitriou, D., Della Valle, M.,  Rau, A.
et al. 2011, \aap, 533, A52, \doi{10.1051/0004-6361/201015887}
% M31 CNe ton-toff diagram
% 10 authors

\bibitem[Y. Hillman et al. (2014)]{hil14}
Hillman, Y., Prialnik, D., Kovetz, A., Shara, M. M., \& Neill, J. D.
2014, \mnras, 437, 1962
\doi{10.1093/mnras/stt2027}
% Nova multiwavelength light curves: predicting UV precursor flashes
% and pre-maximum halts

\bibitem[Y. Hillman et al. (2016)]{hil16}
Hillman, Y., Prialnik, D., Kovetz, A., \& Shara, M. M.
2016, \apj, 819, 168
\doi{10.3847/0004-637X/819/2/168}
% Growing WDs to the Chandrasekhar limit: the parameter space
%of the SD SN Ia channel

\bibitem[R. Hounsell et al. (2010)]{hou10bh}
Hounsell, R., Bode, M. F., Hick, P. P., et al. 2010, \apj, 724, 480,
\doi{10.1088/0004-637X/724/1/480}
%-486.  Exquisite Nova Light Curves from the Solar Mass Ejection Imager (SMEI)

\bibitem[I. Jr., Iben (1982)]{ibe82}
Iben, I. Jr. 1982, \apj, 259, 244, \doi{10.1086/160164}
% hot accreting WDs in the quasi-static approximation

\bibitem[C.A. Iglesias \& F. J. Rogers (1993)]{igl93} Iglesias, C.A., \& Rogers, F.J.
  1993, \apj, 412, 752, \doi{10.1086/172958}
% OPAL

\bibitem[C.A. Iglesias \& F.J. Rogers (1996)]{igl96}
Iglesias, C. A., \& Rogers, F. J. 1996, \apj, 464, 943,
\doi{10.1086/177381}
%-953. Updated Opal Opacities

\bibitem[F. Jablonski \& A. Oliveira (2016)]{job16}
Jablonski, F., \& Oliveira, A. 2016, ATel, 9684, 1
%  Pre-discovery observations of the MASTER OT J010603.18-744715.8 
% Classical Nova in the SMC

\bibitem[M. Kato (1983)]{kat83} Kato, M., 1983, \pasj, 35, 507
% no doi number
% optically thick winds in nova outburst


\bibitem[M. Kato \& I. Hachisu (1994)]{kat94h}
Kato, M., \& Hachisu, I., 1994, \apj, 437, 802, \doi{10.1086/175041}
%Kato, M. \& Hachisu, I. Optically thick winds in nova outbursts.
%{\it Astrophys. J.} {\bf 437}, 802---826 (1994). 

%\bibitem[M. Kato \& I. Hachisu (2011)]{kat11drag}
%Kato, M., \& Hachisu, I., 2011, \apj, 743, 157, 
%\doi{10.1088/0004-637X/743/2/157}
% Effects of a companion star on slow nova outburst -- transition from
% static to wind evolutions

%\bibitem[M. Kato et al. (2009b)]{kat09v838her}
%Kato, M., Hachisu, I., 
%Cassatella, A. 2009, \apj, 704, 1676, \doi{10.1088/0004-637X/704/2/1676}
% V838 Her

\bibitem[M. Kato et al. (2013)]{kat13hh}
 Kato, M., Hachisu, I., Henze, M. 2013, \apj, 779, 19
\doi{10.1088/0004-637X/779/1/19}
% nove in globular clusters   GCnova


\bibitem[M. Kato et al. (2017a)]{kat17palermo}
Kato, M., Hachisu, I.,  \& Saio, H. 2017a, PoS, GOLDEN 2017, 56
in Proceedings of the Palermo
Workshop 2017 on ``The Golden Age of Cataclysmic Variables and
Related Objects - IV'', eds. F. Giovannelli et al. (Trieste: SISSA PoS),
315, 56, \doi{10.22323/1.315.0056}
 (arXiv: 1711.01529)
 https://pos.sissa.it/315/056/pdf
% Palermo conference 2017. Recurrent novae and long term evolution
% of mass accreting WD -- toward the accurate mass retention efficiency

\bibitem[M. Kato et al. (2025)]{kat25v1674her}
Kato, M., Hachisu, I., Saio, H. 2025, \apj, 988,112
% Comprehensive light curve model of the very fast nova V1674 Her
\doi{10.3847/1538-4357/ade231}



%\bibitem[M. Kato et al. (2021)]{kat21sh}
%Kato, M., Saio, H, \& Hachisu, I. 2021, \pasj, 73, 1137,
%\doi{10.1093/pasj/psab0764}
% model V2491 Cyg, H-ignition for massive WD

\bibitem[M. Kato et al. (2022a)]{kat22sha}
Kato, M., Saio, H., \& Hachisu, I. 2022a, \pasj, 74, 1005,
\doi{10.1093/pasj/psac051}
%in press (arXiv:2206.03136)
% Physics of nova outbursts - A theoretical model of
% classical nova outbursts with self-consistent wind mass loss
% 1.0 Mo model

\bibitem[M. Kato et al. (2022b)]{kat22shapjl}
Kato, M., Saio, H, \& Hachisu, I. 2022b, \apjl, 935, L15,
\doi{10.3847/2041-8213/ac85cl}
% X-ray-flash light-curve model for YZ Ret

\bibitem[M. Kato et al. (2022c)]{kat22shc}
Kato, M., Saio, H, \& Hachisu, I. 2022c, Research notes of the AAS, 6, 258,
\doi{10.3847/2515-5172/aca8af}
% X-ray flash for YZ Ret

\bibitem[M. Kato et al. (2024)]{kat24M1213}
Kato, M., Saio, H, \& Hachisu, I. 2024, \pasj,  76, 666,
\doi{10.1093/pasj/pase038}
%%%(arXiv:2404.11237),
% physics of nova outburst , 1.2 and 1.3 WDs

\bibitem[M. Kato et al. (2014)]{kat14shn}
Kato, M., Saio, H., Hachisu, I., Nomoto, K. 2014, \apj, 793, 136
% shortest recurrence periods of novae
%  \& Nomoto, K.

%\bibitem[M. Kato et al. (2016)]{kat16xflash}
% Kato, M., Saio, H., Henze, M. et al. 2016, \apj, 830, 40,
%\doi{10.3847/0004-637X/830/1/40}
% X-ray Flashes in Recurrent Novae: M31N 2008-12a
% and the Implications of the Swift Nondetection
%Kato, Mariko; Saio, Hideyuki; Henze, Martin; et al.
%The Astrophysical Journal, Volume 830,Issue 1, article id.40,12 pp.(2016).
% Pub Date: October 2016 DOI: 10.3847/0004-637X/830/1/40
% Kato, Mariko; Saio, Hideyuki; Henze, Martin; Ness, Jan-Uwe;
% Osborne, Julian P.; Page, Kim L.; Darnley, Matthew J.; Bode, Michael F.;
% Shafter, Allen W.; Hernanz, Margarita; Gehrels, Neil; Kennea, Jamie;
% Hachisu, Izumi


\bibitem[O. K{\"o}nig et al. (2022)]{kon22wa}
K{\"o}nig, O., Wilms, J., Arcodia, R., et al. 2022, \nat, 605, 248,
\doi{10.1038/s41586-022-04635-y}
%% nature, 605, 248    12 May 2022
%% https://doi.org/10.1038/s41586-022-04635-y.
%%YZ Ret


%\bibitem[A. Kovetz (1998)]{kov98}
%Kovetz, A. 1998, \apj, 495, 401, \doi{10.1086/305280}
%-406. The Simulation of Nova Evolution with Optically Thick Winds

\bibitem[J. Krautter et al. (1996)]{kra96}
Krautter, J., \"Ogelman, H., Starrfield, S., Wichmann, R.,
\& Pfeffermann, E. 1996, \apj, 456, 788
\doi{10.1086/176697}
% ROSAT X-ray obs. V1974 Cyg  first X-ray light curve

\bibitem[R. Lico et al. (2024)]{lic24} Lico, R. et al. 2024 \aap, 692,A107
\doi{10.1051/0004-6361/202451364}
% High-resolution imaging of the evolving bipolar outflows in
% RS Oph 2021    radio image from t= 14d

%\bibitem[C.-C. Lin et al. (2022)]{lin22}
%Lin, L, C.-C., Fan, J.-L., Hu, C.-P.,
%Tanaka, J., \& Li, K.-L. 2022, \mnras, 517, L97,
%\doi{10.1093/mnrasl/slac117}
% orbital period and X-ray, gamma-ray in nova her 2021, V1674 Her

%\bibitem[M. Livio (1992)]{liv92}
%Livio, M. 1992, \apj, 393, 516, \doi{10.1086/171524}
%-522. Classical novae and the extragalactic distance scale
%\bibitem[Mallama \& Skillman (1979)]{mal79}
%Mallama, A. D., \& Skillman, D. R. 1979, \pasp, 91, 99,
%\doi{10.1086/130449}
%-100.  Photometry of Nova CYG 1978.


\bibitem[V. Lipunov et al. (2016)]{lip16}
Lipunov, V., Podesta, R., Levato, H., et al. 2016, ATel, 9631, 1 
% Pre-discovery observations of the MASTER OT J010603.18-744715.8 
% Classical Nova in the SMC

\bibitem[X. Ma et al. (2013)]{ma13}
Ma, X., Chen, X., Chen, H-l, Denissenkov, P.A., Han, Z. 2013, \apjl, 778 32,
\doi{10.1088/2041-8205/778/2/L32}
% A Super-Eddington wind scenario for the progenitors of type Ia SNe

\bibitem[P. Martin et al. (2018)]{mar18dj}
Martin, P., Dubus, G., Jean, P., Tatischeff, V., \& Dosne, C. 2018, \aap,
612, A38, \doi{10.1051/0004-6361/201731692}
%Astronomy & Astrophysics, Volume 612, id.A38, 16 pp.
% Gamma-ray emission from internal shocks in novae

\bibitem[J. Maconald (1980)]{mac80}
MacDonald, J. 1980, \mnras, 191, 933, \doi{10.1093/mnras/191.4.933}
% The effect of a binary companion on a nova outburst
% MacDonald, J.
% Monthly Notices of the Royal Astronomical Society, vol. 191,
% June 1980, p. 933-949.
% Pub Date: June 1980 DOI: 10.1093/mnras/191.4.933 

\bibitem[D. B. McLaughlin (1942)]{mcl42}
McLaughlin, D. B. 1942, \apj, 95, 428, \doi{10.1086/144414}
%-436.  Spectral Stages of Novae.

\bibitem[B. D. Metzger et al. (2015)]{met15fv}
Metzger, B. D., Finzell, T., Vurm, I., et al. 2015, \mnras, 450, 2739,
\doi{10.1093/mnras/stv742}
% Gamma-ray novae as probes of relativistic particle acceleration
% at non-relativistic shocks

\bibitem[K. Mukai \& M. Ishida (2001)]{muk01i}
Mukai, K., \& Ishida, M. 2001, \apj, 551, 1024, \doi{10.1086/320220}
%-1030. The Early X-Ray Emission from V382 Velorum (Nova Velorum 1999):
% An Internal Shock Model

\bibitem[K. Mukai \& J. L. Sokoloski (2019)]{muk19s}
Mukai, K., \& Sokoloski, J. L. 2019, Physics Today 72, 11, 38,
\doi{10.1063/PT.3.4341}
%The new science of novae
%Citation: Physics Today 72, 11, 38 (2019); doi: 10.1063/PT.3.4341

\bibitem[U. Munari et al. (2015)]{mun15mm}
Munari, U., Maitan, A., Moretti, S., Tomaselli, S. 2015, NewA, 40, 28,
\doi{10.1016/j.newast.2015.03.008}
% 500 days of Stromgren b, y and narrow-band [OIII], H_alpha
% photometric evolution of gamma-ray Nova Del 2013 (=V339 Del)

\bibitem[U. Munari et al. (2017)]{mun17}
Munari, U., Hambsch, F.-J., \& Frigo, A. 2017, \mnras, 469, 4341
%-4358  (arXiv:1703.09017)
%  Photometric evolution of seven recent novae and the double
%  component characterizing the lightcurve of those emitting in gamma rays

\bibitem[S. Naik et al. (2009)]{naik09}
Naik, S., Banerjee, D. P. K., \& Ashok, N. M. 2009, \mnras, 394, 1551,
\doi{10.1111/j.1365-2966.2009.14421.x}
% Near-infrared observations of the novae V2491 Cygni and V597 Puppis
% Naik, Sachindra search by orcid ; Banerjee, D. P. K. ; Ashok, N. M.
% Monthly Notices of the Royal Astronomical Society, Volume 394,
% Issue 3, pp. 1551-1558.
% Pub Date: April 2009 DOI: 10.1111/j.1365-2966.2009.14421.x 

\bibitem[S. Nakano \& N. N. Samus (2013)]{nak13s}
Nakano, S., \& Samus, N. N. 2013, \iaucirc, 9258, 1
% V339 Delphini = Nova Delphini 2013 = Pnv J20233073+2046041

\bibitem[K. Nariai et al. (1980)]{nar80}
Nariai, K., Nomoto, K., \& Sugimoto, D. 1980, \pasj, 32, 473
% Nova explosion on mass-accreting WDs
% doi number is not found in ADS

\bibitem[J.-U. Ness et al. (2008c)]{nes08c}
Ness, J.-U., et al. 2008c, \iaucirc, 8911, 2
%  V597 Puppis

\bibitem[G. Newsham et al. (2014)]{new14}
Newsham, G., Starrfield, S., \& Timmes, F.X. 2014, ASP conference
Series, Vol.490, Stella Novae: Past and Future Decades,287
% no DOI number
% Evolution of accreting WDs: some of them continue to grow

\bibitem[T. Nelson et al. (2012a)]{nel12a}
Nelson, T., Donato, D., Mukai, K., Sokoloski, J., \& Chomiuk, L. 2012a,
\apj, 748, 43, \doi{10.1088/0004-637X/748/1/43}
%%(16 pp) X-Ray Emission from an Asymmetric Blast Wave
%%   and a Massive White Dwarf in the Gamma-Ray Emitting Nova V407 Cyg

\bibitem[K. Nomoto (1982)]{nom82}
Nomoto, K. 1982, \apj, 253, 798, \doi{10.1086/159682}
%-810. Accretion White Dwarf Models for Type I Supernovae. I.
% Presupernova Evolution and Triggering Mechanisms
% Astrophysical Journal, Part 1, vol. 253, Feb. 15, 1982, p. 798-810.
% Pub Date: February 1982 DOI: 10.1086/159682 

%\bibitem[M. Orio et al. (2022)]{ori22}
%Orio, M., Gendreau, K., Giese, M., et al.
%2022, \apj, 932, 45, \doi{10.103847/1538-4357/ac63be}
% Cal 83, MR Vel, YZ Ret, V1674 Her

\bibitem[A. \"Ozd\"ormez et al. (2018)]{ozd18}
\"Ozd\"ormez, A., Ege, E., G\"uver, T., \& Ak, T. 2018, \mnras,
476, 4162, \doi{10.1093/mnras/sty432}
%  A new catalogue of Galactic novae: investigation of the MMRD
%  relation and spatial distribution

\bibitem[K. L. Page et al. (2022)]{pag22}
Page, K. L., Beardmore, A. P., Osborne, Jp, P.,
Munari, U., Ness, U.-U. et al. 2022, \aap, 514, 1557
% The 2021 outburst of the recurrent nova RS Ophiuchi observation
% in X-rays by the Neil Gehrels Swift Observatory: a comparative study

\bibitem[J. Patterson et al. (2022)]{pat22}
Patterson, J., Enenstein, J, de Miguel, E.,  et al. 2022, \apjl, 940, L56,
\doi{10.3847/2041-8213/ac9ebe}
% V1674 Herculis (2021)

\bibitem[C. Payne-Gaposchkin (1957)]{pay57}
Payne-Gaposchkin, C. 1957, The galactic Novae (Amsterdam: North-Holland)
% no doi

\bibitem[A. Pereira et al. (1999)]{per99}
Pereira, A., di Cicco, D., Vitorino, C., \& Green, D. W. E.
1999, \iaucirc, 7323, 1 
%  Nova Aquilae 1999 No. 2

\bibitem[A.J.S. Pereira et al. (2007)]{per07}
Pereira, A. J. S., McGaha, J. E., Young, J., \& Rhoades, H.
2007, \iaucirc, 8895, 1
%  Nova Puppis 2007

\bibitem[D. Prialnik \& A. Kovetz (1992)]{pri92k}
Prialnik, D., \& Kovetz, A. 1992, \apj, 385, 665, \doi{10.1086/170972}
%  Classical Nova Models with Accretion Heating at Accretion Rates
%  of 10^-9 and 10^-10 Msun per Year

\bibitem[D. Prialnik \& A. Kovetz (1995)]{pri95k}
Prialnik, D., \& Kovetz, A. 1995, \apj, 445, 789, \doi{10.1086/175741}
%  An extended grid of multicycle nova evolution models

\bibitem[C. L. N. Ruggles \& G. T. Bath (1979)]{rug79}
Ruggles, C. L. N. \& Bath, G. T. 1979, \aap, 80, 97
% The structure of optically thick winds in classical novae
% Astronomy and Astrophysics, vol. 80, no. 1, Nov. 1979, p. 97-103.
% Pub Date: November 1979

\bibitem[K. J. Shen \& E. Quataert (2022)]{she22}
Shen, K. J, \& Quataert, E. 2022, \apj, 938, 31,
\doi{10.3847/1538-4357/ac9136}
% Binary interaction dominates mass ejection in CNe

%\bibitem[R. M. Quimby et al. (2024)]{qui24}
%Quimby, R. M., Metzger, B. D., 
%Shen, K.J., et al. 2024, \apj, 977, 17, \doi{10.3847/1538-4357/ad887f}
% Shafter, A. W., Corbett, H., \& Overton, M.
% The rise of nova V1674 Herculis

%\bibitem[G. H. Rieke \& M. J. Lebofsky (1985)]{rie85}
%Rieke, G. H., \& Lebofsky, M. J. 1985, \apj, 288, 618, \doi{10.1086/162827}
%-621, The interstellar extinction law from 1 to 13 microns

%\bibitem[S. Schandl et al. (1997)]{sch97mm}
%Schandl, S., Meyer-Hofmeister, E., \& Meyer, F. 1997, \aap, 318, 73
%Visual light from the eclipsing supersoft X-ray source CAL 87.
%{\it Astron. Astrophys.} {\bf 318}, 73---80 (1997).
% No doi number

\bibitem[B. E. Schaefer (2022a)]{schaefer22a}
Schaefer, B.E., 2022a,\mnras, 517, 3640, \doi{10.1093/mnras/stac2089}
% Comprehensive listing of 156 reliable orbital periods of nova
% including 49 new periods
%  Res Notes AAS, 5, 150;  (arXiv:2207.02932)
%Discovery of 13 New orbital periods for CNe

\bibitem[B. E. Schaefer et al. (2022)]{schaefer22wh}
Schaefer, B. E., Walter, F. M., Hounsell, R., Hillman, Y. 2022, \mnras,
517, 3864, \doi{10.1093/mnras/stac2923}
%  The nova KT Eri is a recurrent nova with a recurrence time-scale of
%  40-50 yr.

\bibitem[B. E. Schaefer (2022b)]{schaefer22b}
%%Comprehensive catalogue of the overall best distances and properties
%% of 402 galactic novae
Schaefer, B. E. 2022b, \mnras, 517, 6150, \doi{10.1093/mnras/stac2900}
%Monthly Notices of the Royal Astronomical Society, Volume 517, Issue 4,
% pp.6150-6169
% Pub Date: December 2022 DOI: 10.1093/mnras/stac2900

\bibitem[G. J. Schwarz et al. (2007)]{sch07}
Schwarz, G. J., Shore, S. N., Starrfield, S., \& Vanlandingham, K. M.
2007, \apj, 657,453, \doi{10.1086/5106611}
% -464 recent V838 Her abundance (van97,sch07)

\bibitem[G. J. Schwarz et al. (2011)]{sch11}
Schwarz, G. J., Ness, J.-U., Osborne, J. P., Page, K. L., Evans, P. A.,
et al. 2011, /apjs, 197, 31,
\doi{10.1088/0067-0049/197/2/31}
% Swift X-ray observations of CNe II. the SSS sample

\bibitem[M. J. Seaton et al. (1994)]{sea94}
Seaton, M. J., Yan, Y., Mihalas, D., Pradhan, A., K., 1994, \mnras, 266, 805,
\doi{10.1093/mnras/266.4.805}
% no DOI number
% Opacities for stellar envelopes
% Seaton, M. J. ; Yan, Y. ; Mihalas, D. ; Pradhan, A. K.
% Monthly Notices of the Royal Astronomical Society, Vol. 266, 805 (1994)
% Pub Date: February 1994 DOI: 10.1093/mnras/266.4.805 

\bibitem[E. M. Sion et al.(1979)]{sio79}
Sion, E. M., Acierno, M.J., \& Tomczyk, S. 1979, \apj, 230, 832,
\doi{10.1086/157143}
%-838  hydrogen shell flashes in massive accreting WDs

\bibitem[K. V. Sokolovsky et al. (2022)]{sok22ll}
Sokolovsky, K. V., Li, K.-L., Lopes de Oliveira, R., et al. 2022, \mnras,
514, 2239, \doi{10.1093/mnras/stac1440}
%% arXiv:2108.03241
% The first nova eruption in a novalike variable: YZ Ret, X-rays and gamma-rays
% Sokolovsky, Kirill V.; Li, Kwan-Lok ; Lopes de Oliveira, Raimundo; et al.
% Monthly Notices of the Royal Astronomical Society,
% Volume 514, Issue 2, pp.2239-2258
% Pub Date: August 2022 DOI: 10.1093/mnras/stac1440

\bibitem[K. V. Sokolovsky et al. (2023)]{sok23}
Sokolovsky, K. V., Johnson, T.J., Buson, S., et al. 2023, \mnras, 521,5453,
\doi{10.1093/mnras/stad887}
% multiwavelength view of shocks in the fastest nova V1674 Her
% X-ray spectrum fit with Nustar. 

\bibitem[W. N. Sparks et al. (1978)]{spa78}
Sparks, W. N., Starrfield, S., \& Truran J. W. 1978, \apj, 220, 1063,
\doi{10.1086/155992}

\bibitem[Starrfield et al. (2009)]{sta09ih}
Starrfield, S., Iliadis, C., Hix, W. R., Timmes, F. X., \& Sparks, W. M.
2009, \apj, 692, 1532, \doi{10.1088/0004-637X/692/2/1532}
%%% The Astrophysical Journal, Volume 692, Issue 2, pp. 1532-1542 (2009).
% The Effects of the pep Nuclear Reaction and Other Improvements
% in the Nuclear Reaction Rate Library on Simulations of
% the Classical Nova Outburst

\bibitem[S. Starrfield et al. (2020)]{sta20}
Starrfield, S., Bose, M., Iliadis, C. et al. 2020, \apj, 895, 70,
\doi{10.3847/1538-4357/ab8d23}
% total 6 authors  Li and other elements. CO mixted at the ignition

\bibitem[A. Tajitsu et al. (2015)]{taj15sn}
Tajitsu, A., Sadakane, K., Naito, H., Arai, A., Aoki, W. 2015,
\nat, 518, 381, \doi{10.1038/nature14161}
% Lithium production in CN V339 Del (nova Delphini 2013)

\bibitem[K. M. Vanlandingham et al. (1997)]{van97}
Vanlandingham, K. M., Starrfield, S., \& Shore, S. N. 1997,  \mnras, 290, 87, 
\doi{10.1093/mnras/290.1.87}
% -98  V693 CrB, also corrected abundance of V838 Her

\bibitem[K. M. Vanlandingham et al. (1996)]{van96}
Vanlandingham, K. M., Starrfield, S., Wagner, R. M., Shore, S. N.,
\& Sonneborn, G.  1996, \mnras, 282, 563
\doi{10.1093/mnras/282.2.563}
% -579 V838 Her abundance determination from IUE spectrum: corrected in
% van97,sch07 summary of E(B-V), detailed explanation of usage of CLOUDY

\bibitem[B. Warner (1995)]{war95}
Warner, B. 1995, Cataclysimic variable stars, Cambridge astrophysics series:28,
(Cambridge University Press: New York

\bibitem[F. M. Walter et al. (2012)]{wal12bt}
Walter, F. M., Battisti, A., Towers, S. E., Bond, H. E.,
\& Stringfellow, G. S. 2012, \pasp, 124, 1057, \doi{10.1086/668404}
%-1072. The Stony Brook/SMARTS Atlas of (mostly) Southern Novae

\bibitem[B. Warner \& P.A. Woudt (2009)]{war09}
Warner, B., \& Woudt, P. A. 2009, \mnras, 397, 979,
\doi{10.1111/j.1365-2966.2009.15006.x}
%-984.  The eclipsing Intermediate Polar V597 Pup (Nova Puppis 2007)

%\bibitem[C. E. Woodward et al. (2021)]{woo21}
% Woodward, C. E., Banerjee D. P.K., 
%Geballe, T.R. et al. 2021, \apjl, 922, L10, \doi{10.3847/2041-8213/ac3518}
% V1674 her, near-IR

\bibitem[C. Wu et al. (2017)]{wu17}
Wu, C., Wang, B., Liu, D., \& Han, Z. 2017, \aap, 604, 31,
\doi{10.1051/0004-6361/201630099}
% mass retention efficiencies of He accretion onto carbon-oxygen WD and type Ia SN

\bibitem[Yamaoka et al. (2009)]{yam09ig}
Yamaoka, H., Itagaki, K., Guido, E. 2009, IAUC, 9098, 1
% Possible nova in eridanus.

%\bibitem[O. Yaron et al. (2005)]{yar05}
%Yaron, O., Prialnik, D., Shara, M. M., \& Kovetz, A. 2005, \apj, 623, 398,
%\doi{10.1086/428435}
% an extended grid of nova models II. add some model to pri95

\end{thebibliography}
\end{document}